\documentclass[multphys,vecphys]{svmult}

\usepackage{makeidx}         
\usepackage{graphicx}        
\usepackage[bottom]{footmisc}

\begin{document}

\title*{Chemical abundances as population tracers}
\author{Poul Erik Nissen}
\institute{Department of Physics and Astronomy, University of Aarhus,
DK-8000 Aarhus C, Denmark. \texttt{pen@phys.au.dk}}
%
%

\maketitle

\begin{abstract}

A discussion of elemental abundance ratios 
as tracers of stellar populations is presented.
The emphasis is on F, G, and K stars, 
because they represent a wide range of ages and have atmospheres 
providing a `fossil' record of the chemical evolution of the
Galaxy.   

Instrumentation and methods to determine chemical abundances in
stellar atmospheres are discussed in Sect. \ref{sect:abundet}.
High-resolution ($R > 20\,000$) spectra
are required to derive
precise abundance ratios, but lower resolution spectra may be
useful in connection with large statistical studies of populations.
Most abundance analyses are based on homogeneous 1D model atmospheres
and the assumption of local thermodynamic equilibrium (LTE), but
recent works have shown that 3D non-LTE corrections can change
the derived trends of abundance ratios as a function of stellar 
metallicity significantly. However, when comparing stars having similar
effective temperatures, surface gravities and metallicities, 3D non-LTE
corrections tend to cancel out. Such a differential approach
is the best way to disentangle stellar populations on the basis
of chemical abundances.

Abundance ratios particularly useful as population tracers
are discussed in Sect. \ref{sect:tracers}, including C/O, 
Na/Fe, Ni/Fe, Ba/Y, Eu/Ba, and $\alpha$/Fe, where $\alpha$ is 
the average abundance of Mg, Si, Ca, and Ti. The nucleosynthesis of
the elements involved occurs on different time-scales in stars and supernovae 
with different masses. This is the main reason that these
abundance ratios can be used as population tracers.

The following sections deal with a discussion of populations in 
the Galactic disk, the bulge, and the halo. Based on abundance
ratios, there is clear evidence for two main populations 
in the disk: an old, thick disk formed on a time-scale of
$\sim \! 10^9$\,years,  and a younger, thin disk formed over a
more extended period. For the bulge, interesting new abundance 
results have been obtained in recent
years, including data from microlensed dwarfs,
but it is too early to draw any robust conclusions about
how and when the bulge formed. For the halo, there is
evidence for the existence of two discrete populations with 
low and high values of $\alpha$/Fe, respectively.
The `low-$\alpha$' population has probably
been accreted from dwarf galaxies, whereas the `high-$\alpha$'
population may consist of ancient disk stars `heated' to halo kinematics 
by merging satellite galaxies. 
Globular clusters stand out from the halo field stars by showing
Na-O and Al-Mg anti-correlations; there is
increasing evidence that they consist of multiple stellar populations. 

\end{abstract}
 

\section{Introduction}
\label{sect:intro}
A population consists of a group of stars with a common origin and history.
Hence, it is of high importance for studies of the formation
and evolution of the Galaxy to detect and describe existing 
Galactic populations. This may be done by analyzing 
distribution functions for stars in space, kinematics, age and chemical composition.
In particular, it is important
to know if the main Galactic components, the disk, the bulge,
and the halo, each consists of a single stellar population or if multiple
populations are needed to fit data for kinematics, ages and 
abundance ratios of stars belonging to these components.

Whereas the original spatial and kinematical distributions of stars in a
population are modified during the dynamical evolution of the 
Galaxy, it is generally assumed that the chemical composition of a
stellar atmosphere provides a 
`fossil' record of the composition of the Galaxy at the time
and the place for the formation of the star. In this connection, F and  G 
main-sequence and subgiant stars are of particular interest, because
they span an age range as long 
as the lifetime of the Galaxy. Furthermore, they have an upper convection
zone that mixes matter in the atmosphere with deeper layers, which tends
to reduce abundance changes induced by diffusion or accretion processes
(see discussion in Sect. 2.5).
On the other hand, the convection zone is not so deep that elements
produced by nuclear reactions in the stellar interior are brought up
to the stellar surface.
Hence, chemical abundances of F and G main-sequence and subgiant stars are
expected to be good tracers of stellar populations. 

Stars with spectral types different from F and G
are also of importance as tracers of Galactic populations.
O, B, and A stars
may be used to probe the present composition of the Galaxy, but 
in some cases the atmospheric composition is affected by diffusion or
accretion processes. K giants are
very useful as a supplement to the F and  G main-sequence stars,
because they can be observed to greater distances. Their space density is,
however, smaller than that of F and G  dwarfs, and care should be taken
because the atmospheric abundances of some elements, e.g. C and N,
may be affected by convective dredge-up of the products of nuclear 
processes in the stellar interior.

The present review deals with the use of stellar abundance ratios to 
disentangle the various stellar populations in the Galaxy. Methods
to determine chemical abundances in stellar atmospheres are 
discussed in Sect. \ref{sect:abundet} 
with emphasis on the high precision that may
be obtained when analyzing stars in a limited region
of the H-R diagram differentially. In Sect. \ref{sect:abundet},
it is also discussed
if element abundances in F and G main-sequence stars are affected
by diffusion or accretion processes. Sect. \ref{sect:tracers}
contains an inventory of abundance
ratios that are particular useful as tracers of stellar populations,
and a discussion of the nucleosynthesis of the involved elements. 
The following sections \ref{sect:disks}, \ref{sect:bulge}, and
\ref{sect:halo} deal with populations associated with
the Galactic disk, the bulge and the halo including globular clusters
and satellite galaxies. Relations between abundance ratios, 
kinematics and ages will be reviewed, and scenarios for the
origin of the various populations will be discussed. 
Finally, Sect. \ref{sect:conclusions} contains conclusions and 
some thoughts about future observing programmes related to
chemical abundances as population tracers. 

The present chapter focus on stars with metallicities in the range
$-3.0 <$ [Fe/H] $< +0.4$, where [Fe/H] is a logarithmic measure
of the ratio between the number of iron and hydrogen atoms in the star
relative to the same ratio in the Sun
\footnote{For two elements X and Y,
[X/Y] =
${\rm log}(N_{\rm X}/N_{\rm Y})_{\rm star}\,\, - \,\,{\rm log}(N_{\rm X}/N_{\rm Y})_{\rm Sun}$,
where $N_{\rm X}$ and $N_{\rm Y}$ are the number densities of the elements.}.
Extremely metal-poor stars with [Fe/H] $< -3.0$
are discussed by Frebel \& Norris (this volume). To some
extent their chemical abundances are related to single supernovae (SNe) events,
whereas a mixture of SNe with a mass distribution determined by the 
initial mass function (IMF)
have produced the elements in more metal-rich stars.

\section{Determination of stellar abundance ratios}
\label{sect:abundet}

\subsection{Observation and reduction of stellar spectra}
\label{sect:obs}

In order to derive precise abundance ratios, high-resolution
($R = \lambda / \Delta \! \lambda > 30\,000$) 
and high signal-to-noise ($S/N > 100$) spectra
should ideally be obtained. For such spectra, it is possible to define
a reliable continuum and measure equivalent widths of weak spectral lines
that have high sensitivity to abundance changes and low sensitivity to
broadening parameters as microturbulence and collisional damping.
Thanks to the installation of efficient echelle spectrographs
in connection with many large and medium-sized telescopes, 
a large number of high-quality  optical ($3700 < \lambda < 9000$\,\AA) spectra for
F, G, and K stars have been obtained during the last couple of decades. 
The infrared spectral region is still lacking behind, but
important abundance results for K giants in the  Galactic bulge 
have been obtained with the Phoenix spectrograph on the Gemini South
telescope (e.g. Mel\'{e}ndez et al. 2008)
and with the ESO VLT cryogenic echelle spectrograph, CRIRES
(Ryde et al. 2010).

Spectra with somewhat lower resolution  ($R \sim 20\,000$) and $S/N \sim 50$
can also be used for determining abundance ratios, and may be
obtained with multi-object spectrographs such as FLAMES at
the ESO VLT.
This has proven to be a very effective way of getting 
abundance data for stars in globular clusters (e.g. Carretta et al. 2009)
and satellite galaxies (see review of Tolstoy et al. 2009). Furthermore, 
abundances of elements that are represented by many lines 
in stellar spectra, such as Fe and the $\alpha$-capture 
elements Mg, Si, Ca, and Ti, 
can be obtained from medium-resolution spectra
($R \sim 5\,000 - 10\,000$). A good example is the determination of stellar 
abundances for
the Sculptor dwarf spheroidal (dSph) galaxy by Kirby et al. (2009)
with the multi-object spectrograph, DEIMOS, at the Keck II telescope.

Even low-resolution ($R \sim 2\,000$) spectra 
are useful for statistical investigations 
of [$\alpha$/Fe]\footnote{Throughout this chapter, $\alpha$ refers to the average
abundance of Mg, Si, Ca, and Ti, i.e. [$\alpha$/Fe] = 
$\frac{1}{4}$ ([Mg/Fe] + [Si/Fe] + [Ca/Fe] + [Ti/Fe])}
in Galactic surveys such 
a the Sloan Digital Sky Survey (SDSS) (Lee et al. 2011).  
Another large survey, the Radial Velocity Experiment (RAVE), which will
deliver medium resolution spectra ($R \simeq 7\,500$) of $\sim 10^6$ stars
in the near-infrared Ca\,{\sc ii}-triplet region (8410 - 8795\,\AA),
has also the potential of supplying [$\alpha$/Fe] with
a decent precision (Boeche et al. 2008).
In the future, the ESA GAIA mission will make it possible to
determine [$\alpha$/Fe] values for a still larger sample of stars
based also on spectra in the Ca\,{\sc ii}-triplet region, but with
a somewhat higher resolution, $R \simeq 11\,500$.

The reduction of raw spectral data should include background and sky subtraction,
flat-field correction, extraction of spectra, and wavelength calibration.
Standard IRAF\footnote{IRAF is distributed by the National Optical
Astronomy Observatories, which are operated by the Association of
Universities for Research in Astronomy, Inc., under cooperative agreement
with the National Science Foundation.} tasks or special software can be used.
Care should be taken to perform a good flatfielding,
including removal of possible interference fringes such that a reliable
continuum can be defined from wavelength regions free of spectral lines.
After normalization of the spectra, equivalent widths (EWs) of weak spectral
lines can be measured by Gaussian fitting of the line profiles. For   
spectral lines having significant line wings (typically $EW > 70$\,m\AA\ 
in F and G main-sequence stars) the fitting should be performed with 
a Voigt profile. The continuum setting and equivalent
width measurements may be
done interactively with the IRAF task {\tt splot} or can be done
automatically (e.g. Sousa et al. 2007).

\begin{figure}
\centering
\includegraphics{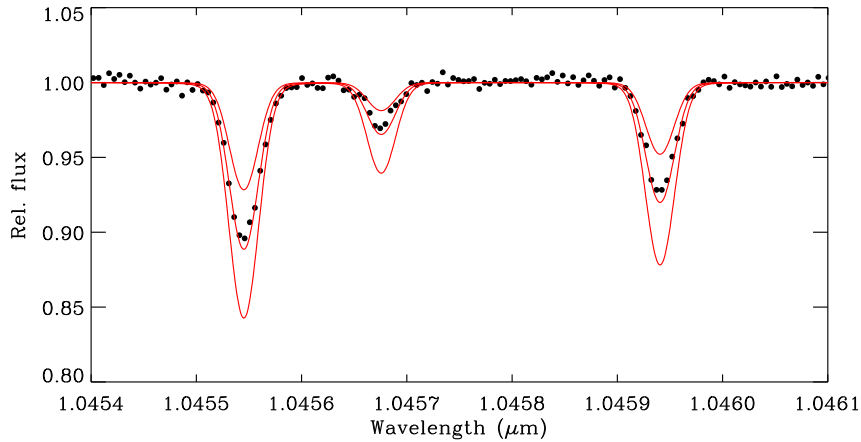}
\caption{The ESO/VLT CRIRES spectrum of the turnoff halo star G\,29-23 
([Fe/H] = $-1.7$) around the near-IR, 1.046\,$\mu$m S\,{\sc i} triplet (dots) 
compared with synthetic LTE model-atmosphere
profiles for three sulphur abundances, 
corresponding to [S/Fe] = 0.0, 0.3 and 0.6, respectively.
As seen, the [S/Fe] = 0.3 case provides a close fit to the observations.
The average S abundance determined from the three lines corresponds
to [S/Fe] = 0.27 (Nissen et al. 2007).}
\label{fig:syn1046.G29-23}
\end{figure}

Instead of using equivalent widths, abundances can be
determined by fitting a
synthetic spectral line profile, calculated for a model atmosphere,
to the observed line profile using the abundance of the element
as a free parameter of the fit (see Fig. \ref{fig:syn1046.G29-23}).
For a line significantly blended by other spectral lines, this
is the only way to derive a reliable abundance. Other lines have
to be used to determine element abundances for the blending
lines and to estimate line broadening parameters associated with
stellar rotation and macroturbulence. Hence, the fitting is an
iterative process that involves several line regions. It 
can be done automatically and may include a determination
of the basic atmospheric parameters, effective
temperature $T_{\rm eff}$, and surface gravity $g$, 
from ratios between selected lines.
A good example of an automatic method is presented by Barklem et al. (2005),
who determine abundances of 22 elements from ESO VLT/UVES spectra
by fitting hundreds of spectral windows containing suitable lines.
The procedure includes the identification of continuum points in the windows,
adjustments of line centers, rejection of lines disturbed by
cosmic-ray hits and
$\chi$-square minimization of the difference between the synthetic
and the observed spectrum. For low- and medium-resolution spectra,
for which individual lines are not resolved and the continuum is
not reached, abundances can be determined by similar methods
(e.g. Lee et al. 2011; Kirby et al. 2009),
or may be based on line-indices that can be calibrated via
high-resolution data or by model-atmosphere calculations.

\subsection{Model atmospheres}
\label{sect:models}

Stellar abundances are normally based on a model-atmosphere analysis of 
the available spectra. In most cases, a plane-parallel, homogeneous
(1D) model is adopted, and it is assumed that the distributions of atoms
over the possible excitation and ionization states are given by
Boltzmann's and Saha's equations. This condition is called `Local Thermodynamic
Equilibrium' (LTE). The temperature structure of the model is derived
from the requirement that the total flux of energy as transported by
radiation and convection should be constant throughout the atmosphere
and given by 
\begin{eqnarray}
F = \sigma \, T_{\rm eff}^4, 
\end{eqnarray}
where $\sigma$ is the Stefan-Boltzmann constant, and $T_{\rm eff}$ 
is the effective temperature of the star. Furthermore, the atmosphere 
is assumed to be in hydrostatic equilibrium and the pressure $P$ as a 
function of optical depth $\tau$ is determined from the equation
\begin{eqnarray}
\frac{dP}{d\tau} = \frac{g}{\kappa_{c}(T,P_e)},
\end{eqnarray}  
where $g$ is the gravity in the stellar atmosphere and $\kappa_{c}$
the continuous absorption coefficient as determined primarily by 
H$^-$ absorption in optical and infrared spectra of F, G, and K stars.
For these cool stars, electrons in the stellar atmosphere mainly
come from the ionization of elements like Mg, Si, and Fe and the
relation between total pressure and electron pressure $P_e$ 
therefore depends on both metallicity and the $\alpha$/Fe ratio.

Details of the construction of 1D stellar models
may be found in textbooks on stellar atmospheres.
The most used grid of models
are the ATLAS9 models of Kurucz (1993) and the Uppsala MARCS models 
(Gustafsson et al. 2008). In both sets of models, convection
is treated in the classical mixing-length approximation.

As reviewed by Asplund (2005), 1D models give only a first approximation
to the temperature structures of stellar atmospheres. The convection
creates an inhomogeneous structure with hot rising granules and cool
downflows. Inhomogeneous (3D) models can be constructed by solving 
the standard equations for conservation of mass, momentum and energy 
in connection with 
the radiative transfer equation for a representative volume of
the stellar atmosphere. The mean temperature structure 
of such 3D models may differ significantly from that of 1D models
especially in the case of metal-poor stars.
Due to the expansion of rising granulation elements and the lack of radiative
heating when the line absorption coefficient is small,
the 3D models have much lower temperature
and electron pressure in the upper layers than classical 1D models in
radiative equilibrium.

\subsection{Abundance analysis}
\label{sect:abunanal}
For a given model atmosphere, the flux $F_{\lambda}$ in an absorption line can
be calculated by solving the transfer equation. Integration over
the line profile relative to the continuum flux $F_c$ then gives
the equivalent width
\begin{eqnarray}
EW = \int{\frac{F_c - F_{\lambda}}{F_c}} d\lambda .
\end{eqnarray}

It is the ratio between the line
and continuous absorption coefficients, $\kappa_l/\kappa_c$,
that determines the line depth and hence the equivalent width. 
For a weak (un-saturated) line, the equivalent width
is approximately proportional
to the abundance ratio $N_{\rm X}/N_{\rm H}$, 
where X is the element corresponding to the line.
For saturated lines, the equivalent width also depends 
on line broadening due to small scale turbulent gas motions. 
In 1D modelling this introduces an additional atmospheric parameter,
the microturbulence, that can be determined from the requirement 
that the same Fe abundance should be derived from
weak and medium-strong  Fe\,{\sc i} lines.  
Strong lines with damping wings are sensitive to the
the value of the collisional damping constant. Clearly, the most
accurate abundances are derived from weak lines if observed with
high resolution and $S/N$. 

The equivalent width of a line also depends on the oscillator strength 
and the populations of
the energy levels corresponding to the line. In LTE,  
Boltzmann's and Saha's equations are used to determine
the population numbers. This may, however, be a poor
approximation as reviewed by Asplund (2005). Instead,
one can use that a stellar atmosphere is in a steady state,
i.e that the population $n_i$ of a level $i$
does not vary in time. This can be expressed as 
\begin{eqnarray}
n_i \sum_{j=1}^{N}{(R_{ij} + C_{ij})} = \sum_{j=1}^{N}{n_j (R_{ji} + C_{ji})},
\end{eqnarray}
where $R$ and $C$ are the transition rates for radiative
and collisional processes, respectively. The summation is
extended over all $N$ levels with $j \neq i$. In such,
so-called non-LTE calculations, 
the population numbers are found  by solving $N$ equations of
the same type as  Eq. (4). In addition,
the transfer equation must be solved, because the radiative 
transition rates depend on the mean intensity of the radiation. 

Departures from LTE can be large and affect derived stellar abundances
very significantly (Asplund 2005). However, in some cases collisional 
transition rates are not well known, and the calculated non-LTE populations 
become rather uncertain. In particular, this is the case for 
inelastic collisions with neutral hydrogen atoms. Often, the recipes
of Drawin (1969)
are adopted, but since these estimates are based on classical
physics, they only provide an order-of-magnitude estimate of the
collisional rates. Hence, a scaling factor $S_{\rm H}$ to the Drawin formula
has to be introduced. It may be calibrated on the basis of solar
spectra by requesting that lines with different excitation potential and
from different ionization stages should provide the same abundance
or one may vary  $S_{\rm H}$ to investigate how the uncertainty of 
collisional rates affects the derived abundances.

Given that non-LTE calculations are sometimes uncertain and that a grid
of 3D models is not yet available, a {\em differential} 1D LTE
analysis is often applied to determine abundance ratios.
For narrow ranges in the basic atmospheric parameters, say $\pm 400$\,K in
$T_{\rm eff}$, $\pm 0.4$\,dex in log$g$, and $\pm 0.5$\,dex in
[Fe/H], one may assume that non-LTE and 3D effects on the
abundances are about the same for all stars.
Hence, precise differential abundances with respect to a standard
star can be derived in LTE. 

For F and G stars with metallicities around
[Fe/H] = 0, the Sun is an obvious
choice as a standard, and logarithmic abundance ratios with 
respect to the Sun, like [Mg/Fe], can be derived from the same
lines in the spectra of the stars and the Sun. At lower metallicities, bright
stars with well known atmospheric parameters can be chosen as standards.
This method has the additional advantage that the oscillator strength of a 
line cancels out so that its error plays no role.
Such differential abundance ratios can be 
determined to a precision of about $\pm 0.03$\,dex (e.g. Neves et al. 2009;
Nissen \& Schuster 2010). When using chemical abundances to
trace stellar population, it is just these very precise differential
abundance ratios at a given metallicity that are needed. Trends of
abundance ratios as a function of [Fe/H] derived under the LTE assumption are, 
on the other hand, less accurate, because non-LTE and 3D effects
change with metallicity.

\subsection{Determination of atmospheric parameters for F, G, and K stars}
\label{sect:param}

In order to determine precise abundance ratios,
reliable values of the stellar atmospheric parameters,
$T_{\rm eff}$, $g$, and [Fe/H], must be determined. Some abundance 
ratios like [Mg/Fe] determined from neutral atomic lines are
fairly insensitive to errors in the atmospheric parameters,
but other ratios like [O/Fe] with the oxygen abundance determined 
from the high excitation O\,{\sc i} triplet or from OH lines
depend critically on the adopted values for $T_{\rm eff}$ and $g$.

The effective temperature of a late-type star can be determined from
a colour index, e.g.  $V\!-\!K$, calibrated in terms of $T_{\rm eff}$
by the infrared flux method. Two recent
implementations of this method (Gonz\'{a}lez Hern\'{a}ndez 
\& Bonifacio 2009; Casagrande et al. 2010) give consistent
calibrations of $V - K$. In the case of nearby stars for which
colours are not affected by interstellar reddening, $T_{\rm eff}$
can be determined to an accuracy of the order of $\pm 50$\,K. For more
distant stars, the reddening is, however, a problem and $T_{\rm eff}$
is better determined spectroscopically, e.g. from the wings of
Balmer lines or from the requirement that [Fe/H]
derived from Fe\,{\sc i} lines should be independent of the
excitation potential of the lines.  In this way, differential values
of $T_{\rm eff}$  can be determined to a precision of $\pm 25$\,K
(Nissen 2008).

The best way to determine the stellar surface gravity 
\begin{eqnarray}
g = G  \frac{\cal{M}}{R^2}
\end{eqnarray}
is to estimate the mass $\cal{M}$ from stellar evolutionary tracks
and the radius $R$ from the basic relation 
$L \propto R^{2} \, T_{\rm eff} ^{4}$,
where $L$ is the luminosity of the star. This leads to the following
expression for the gravity of a star relative to that of the
Sun (log$g_{\odot} = 4.44$ in the cgs system)

\begin{eqnarray}
\log \frac{g}{g_{\odot}}  =  \log \frac{\cal{M}}{\cal{M}_{\odot}} +
4 \log \frac{T_{\rm eff}}{T_{\rm eff,\odot}} +
0.4 (M_{\rm bol} - M_{{\rm bol},\odot}),
\end{eqnarray}
where $M_{bol}$ is the absolute bolometric magnitude, which can be
determined from the apparent magnitude if the distance to the star
is known. 

This method of determining surface gravities works well for nearby stars
for which distances are accurately known from Hipparcos parallaxes.
For more distant late-type stars, the gravity can be determined
spectroscopically from the difference in [Fe/H] derived from neutral and
ionized iron lines. Fe\,{\sc i} lines change very little with $g$,
whereas Fe\,{\sc ii} lines change significantly.  Departures from LTE 
in the ionization
equilibrium of Fe should, however, be taken into account. This may be
done by requiring that the difference
[Fe/H](Fe\,{\sc ii})\,$-$\,[Fe/H](Fe\,{\sc i})
has the same value as in the case of a standard star with a surface 
gravity that is accurately determined from Eq. (6).
In this way differential values of log\,$g$
can be determined to a precision of about $\pm 0.05$\,dex
(Nissen 2008).

Due to the non-LTE effects on the ionization balance of Fe,
[Fe/H] should be determined from
Fe\,{\sc ii} lines, because they represent the dominating ionization
stage of iron. If LTE is assumed, [Fe/H] derived from Fe\,{\sc i}
lines turns out to be 0.1 to 0.2 dex lower than [Fe/H] derived 
from Fe\,{\sc ii} lines in the case of metal-poor F, G, and K stars.
It is likely that this problem is due to a higher degree of 
Fe\,{\sc i} ionization than predicted by the Saha equation
(Mashonkina et al. 2011). 

\subsection{Diffusion and dust-gas separation of elements}
\label{sect:diffusion}
As mentioned in Sect. \ref{sect:intro}, it is generally assumed that
the atmosphere of a late-type star with an upper convection
zone has retained
a `fossil' record of the composition of the Galaxy at the time
and the place for the formation of the star.  A high-resolution study of
the metal-poor ([Fe/H]\,$\sim -2$) globular cluster NGC\,6397 by 
Korn et al. (2007) indicates, however, that the abundances of
Mg, Ca, Ti, and Fe in main-sequence turnoff 
stars are about 0.12\,dex (30\,\%) lower than the abundances of  these
elements in K giants. This may be explained by downward diffusion 
of the elements at the bottom of the convection zone for turnoff stars.
The elements are depleted by about
the same factor, so the effect of diffusion on abundance ratios is
less than 0.05\,dex. 

For the solar atmosphere, the depletion by diffusion of 
elements heavier than boron is predicted to have 
been about 0.04\,dex (Turcotte \& Wimmer-Schweingruber 2002) and 
the effect of diffusion on abundance ratios is negligible. 
This is confirmed by
the good agreement of abundance ratios for non-volatile elements
in the solar atmosphere and in the most primitive meteorites,
the carbonaceous chondrites (Asplund et al. 2009). A very precise
study of solar `twin' stars (i.e. stars having nearly the same $T_{\rm eff}$,
$g$, and [Fe/H] as the Sun) by Mel\'{e}ndez et al. (2009) shows,
on the other hand, that the Sun has a higher abundance ratio of volatile elements
(C, N, O, S, and Zn) with respect to Fe than the large majority of 
twin stars. The deviation is about 0.05\,dex. As suggested by the authors,
this may be explained by selective accretion of refractory elements, 
including iron, on dust particles in the proto-solar disk. Thus, some
fraction of the refractory elements may end up in terrestrial planets. 
If true, abundance ratios like [O/Fe] and [S/Fe] can deviate by
$\sim \! 0.05$\,dex from the original ratio in the interstellar cloud that
formed the star depending on whether the star is with or without 
terrestrial planets.

It is concluded that the effects of diffusion may change some abundance
ratios by up to 0.05\,dex for metal-poor stars with relatively thin 
convection zones. In the case of disk stars, the abundances of volatile
elements relative to iron could be affected by dust-gas separation in
connection with star and planet formation by $\sim\!0.05$\,dex. For
refractory elements there are no indications of differences in 
abundance rations between stars with and without detected planets
(Neves et al. 2009).

\section{Elements used as stellar population tracers}
\label{sect:tracers}

This section presents a discussion of some abundance ratios that 
have been used as tracers of stellar populations.
In several cases, the nucleosynthesis and chemical evolution of
the corresponding elements are not well understood; nevertheless,
the abundance ratios have proven to be very useful in disentangling
stellar populations. In addition, the observed differences and trends
provide important constraints on SNe modeling and theories 
of Galactic chemical evolution. 

\subsection{Carbon and oxygen}
\label{sect:CO}

Abundances of C and O can be determined from spectral lines
corresponding to forbidden transitions between low excitation states
and allowed transitions between high-excitation states for neutral
atoms. In addition, molecular CH and OH lines 
in the blue-UV and infrared spectral regions can be applied. 

The most reliable C and O abundances are derived from the forbidden 
[C\,{\sc i}] $\lambda 8727$ and [O\,{\sc i}] $\lambda 6300$ lines, provided
that these rather weak lines are measured with sufficient spectral resolution
and $S/N$. Both lines are blended; [O\,{\sc i}] by a
Ni\,{\sc i} line and [C\,{\sc i}] by a weak Fe\,{\sc i} line. 
These blends must be
taken into account when deriving abundances. A strong collisional
coupling with the ground states ensures that the [C\,{\sc i}] and
[O\,{\sc i}] lines are formed in LTE. The correction for 3D effects is
small for the Sun (Asplund 2005) but increases with decreasing metallicity
and may reach as much as $-0.2$\,dex in turnoff stars with [Fe/H] = $-2$
(Nissen et al. 2002). 

The [C\,{\sc i}] line is too weak to be a useful abundance indicator for 
metal-poor ([Fe/H]\,$< -1$) dwarf and subgiant stars. In giants, 
carbon abundances are changed by dredge-up of gas affected by CN-cycle
hydrogen burning. The [O\,{\sc i}] line, on the other hand,
can be used to derive oxygen abundances
in dwarfs and subgiants down to [Fe/H]\,$\sim -2$ and in giants down to
metallicities around $-3$. Alternatively, carbon and oxygen abundances 
in halo stars can be derived from high-excitation atomic lines
(the C\,{\sc i} lines around 9100\,\AA\ and the O\,{\sc i} triplet at 
7774\,\AA ) but in both cases the non-LTE effects are uncertain
(Fabbian et al. 2009). CH and OH lines can be used to derive
carbon and oxygen abundances even at extremely low metallicities,
but they are very sensitive to $T_{\rm eff}$ and large 3D corrections
should be applied (Asplund 2005). Due to these problems, C and O
abundances in halo stars are quite uncertain. Carbon seems to follow
iron, i.e. [C/Fe] $\simeq 0$ from [Fe/H] = 0 to $-3$ (Bensby \& Feltzing
2006; Fabbian et al. 2009). [O/Fe] raises from zero to about +0.5\,dex,
when the metallicity of disk stars decreases from [Fe/H]\,=\,0 to $-1$
and then stays approximately constant at 
[O/Fe] $\simeq +0.5$ down to [Fe/H] $\simeq -2$ (Nissen et al. 2002). 
According to Cayrel et al. (2004), who derived oxygen abundances
in giants from the [O\,{\sc i}] line, the constant level of [O/Fe] continues
all the way down to [Fe/H] $\simeq -3.5$, if one assumes that 
3D corrections are the same as in metal-poor dwarf stars.

\begin{figure}
\centering
\includegraphics{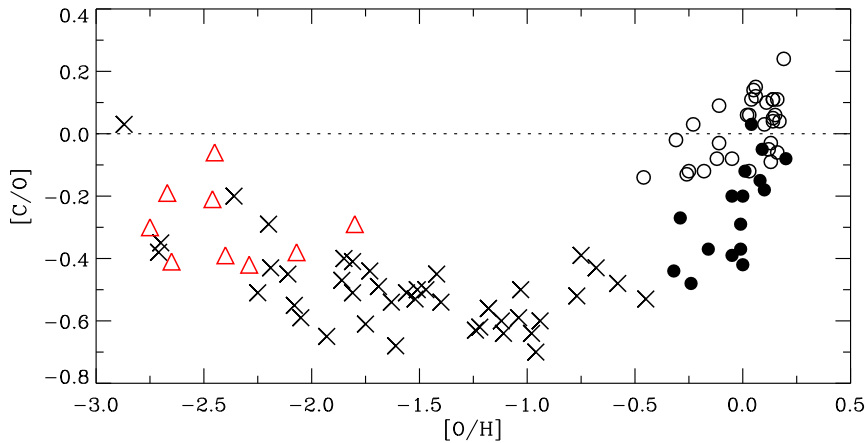}
\caption{The carbon-to-oxygen ratio as a function
of the oxygen abundance. Open circles refer to thin-disk
and filled circles to thick-disk stars with data adopted from
Bensby \& Feltzing (2006). Crosses are halo stars from
Fabbian et al. (2009), and (red) triangles show DLA data from
Cooke et al. (2011).}
\label{fig:CO-O.all}
\end{figure}

Although the trends of [C/Fe] and [O/Fe] as a function of [Fe/H]
are somewhat uncertain due to non-LTE and 3D effects, the ratio
between the abundances of C and O is more immune to these problems.
This stems from the fact that the forbidden [C\,{\sc i}] and 
[O\,{\sc i}] lines have about the same dependence of temperature and
pressure. Hence, the derived [C/O] is insensitive to 3D effects.
The same is the case for [C/O] derived from high-excitation atomic
lines. Furthermore, as shown by Fabbian et al. (2009),
the non-LTE corrections of C and O abundances derived from
C\,{\sc i} and O\,{\sc i} lines tend to cancel, so that the
trend of [C/O] vs. [O/H] is fairly independent of the choice of
the hydrogen collision parameter, $S_{\rm H}$ (see Sect. 2.3).
Fig. \ref{fig:CO-O.all} shows this trend for $S_{\rm H} = 1$,
with the abundances determined from forbidden lines for disk stars
and from high-excitation atomic lines for the halo stars.

Carbon is synthesized in stellar interiors by the triple-$\alpha$ process,
but it is unclear which objects are the main contributors to the
chemical evolution of carbon in the Galaxy. Type II SNe, Wolf-Rayet
stars, intermediate- and low-mass stars in the planetary nebula phase,
and stars at the end of the giant phase have been suggested. 
Oxygen, on the other
hand, seem to be produced exclusively by $\alpha$-capture on C
in short-lived massive stars
and is dispersed to the interstellar medium by type II SNe. Hence, the
C/O ratio has the potential of being a good tracer of stellar populations.
The separation between thin- and thick-disk stars in Fig. \ref{fig:CO-O.all}
shows that this is indeed the case.

The approximately constant [C/O]\,$\simeq -0.5$ for halo stars 
with [O/H] between $-2.0$ and $-0.5$ probably corresponds to the
C/O yield ratio for massive stars. The increase in [C/O] for 
the thick-disk stars may be due to metallicity
dependent winds from Wolf-Rayet stars, whereas 
the delayed production of carbon by low- and intermediate-mass stars can
explain the higher [C/O] in thin-disk stars.
The upturn of [C/O] at the lowest values of [O/H], which is
also found for distant damped Lyman-$\alpha$ (DLA) galaxies 
(Cooke et al. 2011), could be due to enhanced carbon production
by massive first generation stars with extremely high
rotation velocities (Chiappini et al. 2006).

\subsection{Intermediate-mass elements}
\label{sect:intermediate}

The even-$Z$ elements, Mg, Si, S, Ca, and Ti are mainly produced by 
successive capture of $\alpha$-particles in connection with
carbon, oxygen and neon burning in massive stars 
and dispersed into the interstellar medium by type II supernovae 
explosions on a time-scale of $\sim 10^7$ years. Iron is also produced
by SNe II, but the bulk of Fe comes from type Ia SNe on a
much longer time-scale ($\sim 10^9$ years). Hence, the ratio
between the abundance of an $\alpha$-capture element and iron
in a star depends on how long the star-formation process had
proceeded before the star was formed. In this way, [$\alpha$/Fe]
becomes an important tracer of populations. 

Mg, Si, Ca, and Ti abundances can be determined from several weak
atomic lines in the optical spectra of late-type stars, whereas sulphur
is more difficult  as discussed below.
Traditionally, [$\alpha$/Fe] is therefore defined as the average value of
[Mg/Fe], [Si/Fe], [Ca/Fe], and [Ti/Fe] (see Sect. 2.1).
As discussed by Asplund (2005), departures from
LTE have some effects on the metallicity trends of
[$\alpha$/Fe] but probably not more than 0.1\,dex, when weak lines
are used. If the abundances are based on neutral lines,
[$\alpha$/Fe] is quite insensitive to
$T_{\rm eff}$ and 3D effects, although high excitation lines are
to be preferred (Asplund 2005, Fig. 8)

The trend of [$\alpha$/Fe] with metallicity is characterized by
an increase of about 0.3\,dex when [Fe/H] decreases from 0 to $-1$.
Below [Fe/H] = $-1$, [$\alpha$/Fe] is distributed around a
plateau at 0.3\,dex, but as discussed later, there are very
significant differences in [$\alpha$/Fe] at a given metallicity 
related to stellar populations both for disk and halo stars.

Sulphur is an $\alpha$-capture element, and [S/Fe] is therefore expected
to show a plateau-like behaviour for halo stars, but very high
values [S/Fe] $\sim 0.8$ have been claimed at the lowest metallicities
(Israelian \& Rebolo 2001). 
These values may, however, be spurious due
to the difficulty of measuring the very weak S\,{\sc i} line at
8694.6\,\AA . On the basis of stronger S\,{\sc i} lines at 
9212.9 and 9237.5\,\AA\ measured with UVES and corrected for
telluric absorption lines, Nissen et al. (2007) find a plateau-like
behaviour of [S/Fe] as shown in Fig. 3. Non-LTE corrections
from Takeda et al. (2005) corresponding to $S_{\rm H}$ = 1 were
included for sulphur, and the iron abundances were derived from
Fe\,{\sc ii} lines with negligible non-LTE effects 
(Mashonkina et al. 2011) . This result is supported
by CRIRES observations of the near-IR S\,{\sc i} triplet
(see Fig. \ref{fig:syn1046.G29-23}). Further
studies of sulphur in Galactic stars would be important, especially
because S is a volatile element that is undepleted onto dust. As
such it can be used to measure the
$\alpha$-enhancement of DLA galaxies.

\begin{figure}
\centering
\includegraphics{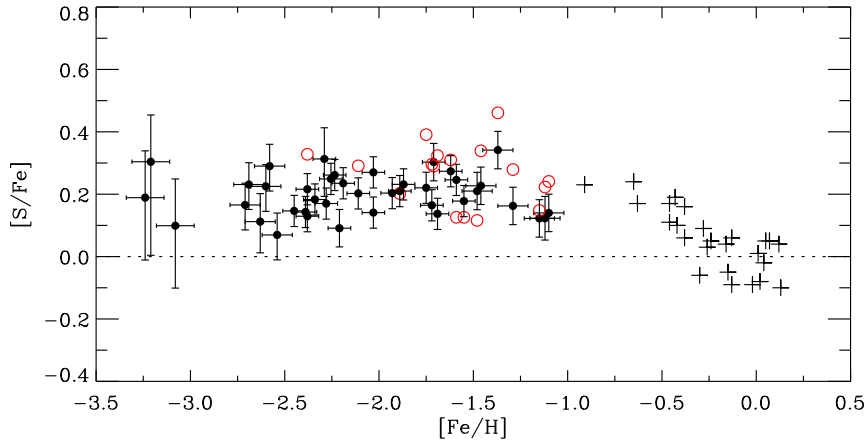}
\caption{[S/Fe] as a function of [Fe/H].
Plus signs refer to data for disk stars from Chen et al. (2002) and circles 
to halo stars from Nissen et al. (2007). 
Filled circles with error bars are based on S abundances derived from
the $\lambda \lambda 9212.9, 9237.5$ S\,{\sc i} lines, whereas
open (red) circles show data determined from
the weak $\lambda 8694.6$ S\,{\sc i} line.}
\label{fig:SFe-Fe.NLTE}
\end{figure}

In addition to the $\alpha$-capture elements, Na is an interesting
and sensitive tracer of stellar populations. It is thought to
be made during carbon and neon burning in massive stars 
and is expelled by type II SNe together with the $\alpha$-elements.
The amount of Na made is, however, controlled
by the neutron excess, which depends on the initial heavy element abundance in
the star (Arnett 1971). This is probably the explanation of the
fact that [Na/Mg] in halo stars correlates with [Mg/H]
with a slope of about 0.5 (Nissen \& Schuster 1997; Gehren et al. 2004).

Na abundances are best determined from the relatively weak Na\,{\sc i} doublets
$\lambda \lambda 5682.6, 5688.2$ and $\lambda \lambda 6154.2, 6160.7$,
for which non-LTE and 3D corrections are rather small (Asplund 2005).
At low metallicities, [Fe/H] $< -2.0$, these lines are too faint to be
measured, and Na abundances are derived from the Na\,{\sc i}\,D 
$\lambda \lambda 5890.0, 5895.9$ resonance lines. For this doublet, the non-LTE
correction is large and reaches $-0.4$\,dex for extremely metal-poor stars
(Gehren et al. 2004).
Furthermore, the use of these lines is sometimes complicated by overlapping
interstellar Na\,{\sc i}\,D lines and telluric H$_2$O lines.

In addition to Na production in massive stars, sodium can also be 
made in hydrogen-burning shells of intermediate- and low-mass stars
via the CNO and Ne-Na cycles. This is probably the explanation of
the Na-O anti-correlation in globular clusters
as further discussed in Sect. \ref{sect:globclusters}.

\subsection{The iron-peak elements}
\label{sect:ironpeak}

Among the iron-peak elements -  
Cr to Zn - the even-$Z$ elements Cr, Fe, and Ni
are represented by many lines in the spectra of late-type stars, which makes
it possible to determine very precise abundance ratios, [Cr/Fe]
and [Ni/Fe]. 

The ratio between Cr and Fe abundances is found to be the same as in the Sun,
i.e. [Cr/Fe] $\simeq 0$, for all Galactic populations with [Fe/H] $> -2$.
Below this metallicity, McWilliam et al. (1995) and Cayrel et al. (2004)
have found a smooth decrease of [Cr/Fe] to $\sim -0.5$ at [Fe/H] = $-4$
based on an LTE study of Cr\,{\sc i} lines in very metal-poor giants. 
This is surprising, because Cr and Fe are predicted
to be synthesized in constant ratios by explosive silicon burning in both 
type II and Ia SNe. The problem seems to have been solved by
Bergemann \& Cescutti (2010), who find that the derived decline of [Cr/Fe] is an
artifact caused by the neglect of non-LTE effects. They
obtain a very satisfactory agreement between non-LTE Cr abundances derived from
Cr\,{\sc i} and  Cr\,{\sc ii} lines when using surface gravities based on Hipparcos
parallaxes.   

For a long time, the Ni/Fe ratio was also thought to be solar in all
stars. This is clearly the case for disk population stars 
(e.g. Chen et al. 2000), but Nissen \& Schuster 
(1997, 2010) have shown that Ni is underabundant relative to Fe, i.e. 
[Ni/Fe] $\sim -0.1$ to $-0.2$, in some halo stars with [Fe/H] $> -1.4$.
These stars have also low [$\alpha$/Fe] and [Na/Fe] values,
and a tight correlation between 
[Ni/Fe] and [Na/Fe] is present (see Sect. \ref{sect:twohalopop}). 
Similar stars are found in dwarf galaxies (Tolstoy et al. 2009).
The reason for this correlation 
is probably that the supernovae yields of $^{23}$Na and $^{58}$Ni depend
on the neutron excess (see the detailed discussion by Venn et al. 2004).

The determination of abundances of the odd-$Z$ elements Mn, Co, and Cu
is more difficult than in the case of Cr, Fe, and Ni.
There are fewer lines available and hyper-fine-structure
(HFS) splitting has to be taken into account. Furthermore, non-LTE
effects seem to be very significant. Positive corrections of
both [Mn/Fe] (Bergemann \& Gehren 2008) and [Co/Fe] 
(Bergemann et al. 2010) are obtained, although
the size of these corrections depends somewhat
on the adopted $S_{\rm H}$ parameter
for hydrogen collisions. This means that the strong decline of [Mn/Fe]
as a function of decreasing [Fe/H] found from an LTE analysis of Mn\,{\sc i}
lines by e.g. Reddy et al. (2006) and Neves et al. (2009) has to be 
modified to a more shallow trend. For Co the non-LTE study
of Bergemann et al. (2010) leads to surprisingly large over-abundances
of cobalt with respect to iron for halo stars, a result that is 
in disagreement with expectations from presently calculated supernovae
yields.

\begin{figure}
\centering
\includegraphics{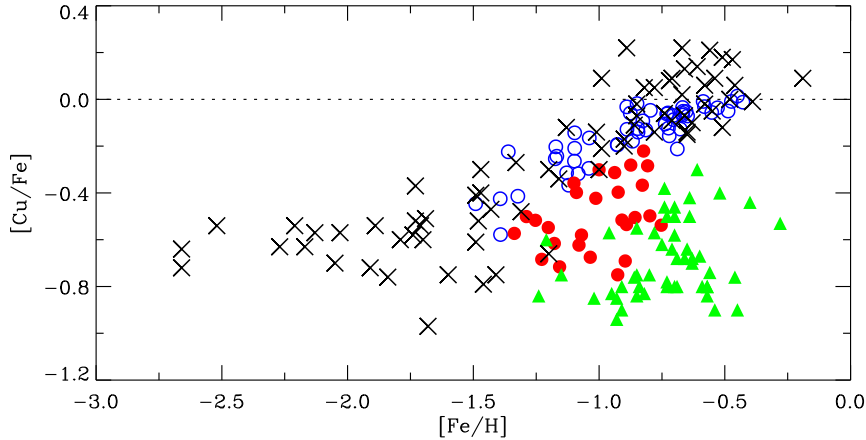}
\caption{[Cu/Fe] as a function of [Fe/H]. Stars studied by
Mishenina et al. (2002) are indicated by crosses. Open (blue) circles refer
to high-$\alpha$ stars and filled (red) circles to low-$\alpha$ halo stars
from Nissen \& Schuster (2011). Filled (green) triangles are data
for giant stars in the inner part of the Large Magellanic Cloud adopted
from Pomp\'{e}ia et al. (2008).
}
\label{fig:Cu-Fe}
\end{figure}

Copper is a very interesting element as a tracer of
metal-poor stellar populations. Abundances can be determined from
the Cu\,{\sc i} lines at 5105.5, 5218.2 and 5782.1\,\AA . They 
are affected by HFS splitting to different degrees.
Non-LTE calculations are not yet available, but LTE data show
that [Cu/Fe] $\simeq 0$ for disk stars with no significant difference
between the thin and the thick disk (Reddy et al. 2006).
At [Fe/H] $\sim -1$, [Cu/Fe] starts to decline steeply and 
reaches a plateau of [Cu/Fe] $\sim -0.6$ below [Fe/H]\,$\simeq -1.6$
(Mishenina et al. 2002)
as shown in Fig. \ref{fig:Cu-Fe}. Low-$\alpha$ halo stars 
deviate, however, from this trend with½ a [Cu/Fe] deficiency
of 0.2 to 0.5\,dex (Nissen \& Schuster 2011). The same is the case
for the more metal-rich part of stars in the globular cluster
$\omega$ Centauri (Cunha et al. 2002),
and stars belonging to the Sagittarius dSph galaxy
(Sbordone et al. 2007) and the Large Magellanic Cloud
(Pomp\'{e}ia et al. 2008) as shown in Fig. \ref{fig:Cu-Fe}.
These [Cu/Fe] data provide important constraints
on the nucleosynthesis of Cu. Romano \& Matteucci (2007)
suggest that copper is initially made by explosive nucleosynthesis in
type II SNe and later by a metallicity dependent neutron-capture process
(the weak $s$-process) in massive stars.

Zinc abundances can be determined from the $\lambda \lambda 4722.2, 4810.5$
Zn\,{\sc i} lines. Non-LTE and 3D corrections are modest, i.e. of the order of
+0.1\,dex for metal-poor stars (Nissen et al. 2007). It is
sometimes assumed that [Zn/Fe] = 0, which means that Zn can be used as a
proxy of Fe in determining the metallicity of interstellar gas, e.g.
in DLA systems; as a volatile element Zn is not depleted onto dust like Fe.
Newer studies show, however, that [Zn/Fe] reaches +0.15\,dex in metal-poor
disk stars with perhaps a small separation between thin- and thick-disk
stars (Bensby et al. 2005). For halo stars, [Zn/Fe]
increases from zero at [Fe/H] = $-1$ to [Zn/Fe] $\simeq 0.2$ at 
[Fe/H] = $-2.5$ (Nissen et al. 2007), and at still lower metallicities
[Zn/Fe] increases steeply to [Zn/Fe] $\simeq 0.5$ at [Fe/H] = $-3.5$
(Cayrel et al. 2004, Nissen et al. 2007). Hence the behaviour of zinc
is complicated, and the nucleosynthesis of this element is not well
understood. Several ways of producing Zn have been suggested:
the weak $s$-process in massive stars, explosive Si burning in type II and
type Ia SNe, and the main $s$-process in low- and intermediate-mass stars.

\subsection{The neutron capture elements}
\label{sect:ncapture}
Among the heaviest elements, Y, Ba, and
Eu are of particular interest as tracers of stellar populations. 
They are made by neutron capture processes, which are
traditionally divided into the slow $s$-process and the 
rapid $r$-process. In the $s$-process, neutrons
are added on a long time-scale compared to that of $\beta$ decays
so that nuclides of the $\beta$-stability valley are built up.
In the $r$-process, neutrons are added so fast that nuclides 
on the neutron-rich side of the stability valley are made.
The $s$-process is divided into the main $s$-process that 
occurs in low- and intermediate-mass asymptotic giant branch (AGB) stars
and the weak $s$-process occurring in massive stars. The $r$-process
is not well understood, but it is thought to occur in connection with
type II SNe explosions.

\begin{figure}
\centering
\includegraphics{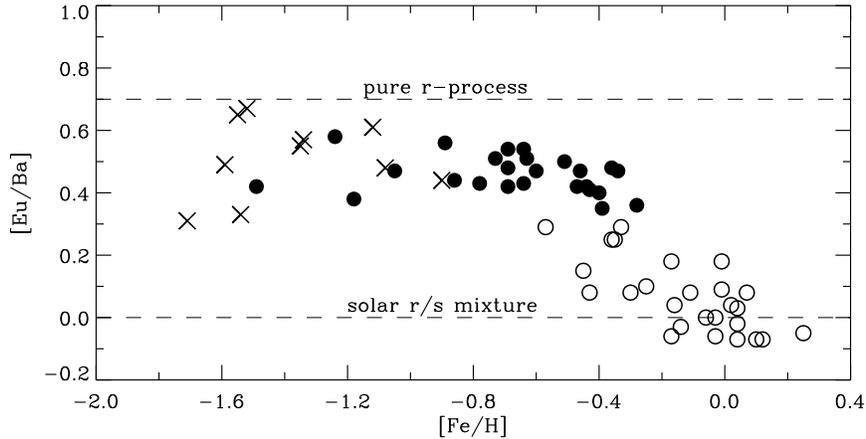}
\caption{[Eu/Ba] as a function of [Fe/H] with data adopted from
Mashonkina et al. (2003). Crosses refer to stars with
halo kinematics, filled circles to thick-disk stars and
open circles to thin-disk stars.}
\label{fig:Eu-Ba}
\end{figure}

The relative contribution of the $s$- and the $r$-process
to heavy elements in the solar system has been determined by
Arlandini et al. (1999). Barium is called an $s$-process 
element, because 81\% of the solar Ba is due to the $s$-process.
Europium, on the other hand, is an  $r$-process element, because
94\% in the Sun originates from the $r$-process. In metal-poor stars,
for which only massive stars have contributed to the nucleosynthesis
of the elements, both Ba and Eu are, however, made by the $r$-process,
provided that the contribution from the weak $s$-process is
negligible. In such metal-poor stars, one would expect to find
an $r$-process ratio $N_{\rm Eu}/N_{\rm Ba} \simeq 5$ between the
abundances of europium and barium corresponding to [Eu/Ba] $\simeq 0.7$. 
These considerations suggest that [Eu/Ba] may be a useful tracer
of stellar populations. As seen from Fig. \ref{fig:Eu-Ba}, this is
confirmed by Mashonkina et al. (2003).

Ba abundances can be determined from subordinate Ba\,{\sc ii} lines
at 5853.7, 6141.7, and 6496.9\,\AA . The stronger
resonance line at 4554.0\,\AA\ may also be used, but the analysis
of this line is complicated by the presence of isotopic and 
HFS splitting. Europium abundances are primarily obtained from the 
Eu\,{\sc ii} line at 4129.7\,\AA , which has to be analyzed by 
spectrum synthesis techniques, because the line is strongly broadened 
by isotopic and HFS splitting.

The barium-to-yttrium ratio is another interesting tracer of stellar populations.
Yttrium ($Z =39$) belongs to the first peak of $s$-process elements
around the neutron magic number $N = 50$, whereas barium ($Z =56$)
is at the second peak around $N = 82$. The ratio Ba/Y
(sometimes called heavy-$s$ to light-$s$, hs/ls) depends on the 
the neutron flux per seed nuclei, and is predicted to be high for
metal-poor, low-mass AGB stars (Busso et al. 1999). As pointed
out by Venn et al. (2004), the Ba/Y ratio in stars belonging to
dSph satellite galaxies is much higher than Ba/Y in Galactic halo stars
with differences on the order of 0.6\,dex in
the metallicity range $-2 <$ [Fe/H] $< -1$. Similar large offsets
are found for stars with [Fe/H] around $-1$ in the globular cluster 
$\omega$\,Cen (Smith et al. 2000) and for stars in the LMC
(Pomp\'{e}ia et al. 2008).
According to Fenner et al. (2006), this indicates that the chemical
evolution of these systems has been so slow that winds from
low-mass AGB stars have started to enrich the interstellar medium
with $s$-process elements at a metallicity around [Fe/H] $\sim -2$.

\newpage

\section{The Galactic disk}
\label{sect:disks}

\subsection{The thick and the thin disk}
\label{thinthick}

A long-standing problem in studies of Galactic structure and evolution
has been the possible existence of a population of stars having
kinematics, ages, and chemical abundances in between the 
characteristic values for the halo and the disk.

On the basis of a large program of
$uvby$-$\beta$ photometry of F and early G type main-sequence stars
within 100\,pc, Str\"{o}mgren (1987) concluded that an
{\em intermediate Population II} does exist. [Fe/H] was determined 
from a metallicity index $m_1$ = $(v-b) - (b-y)$, which is sensitive to
the line blanketing in the $v$-band, and age was derived
from the position of a star in the $c_1$ - $(b-y)$ diagram, where
$c_1$ = $(u-v) - (v-b)$ is a measure of the Balmer discontinuity
at 3650\,\AA\ and hence of the surface gravity of the star. 
The colour index $(b-y)$ is used as a measure of $T_{\rm eff}$.
After discussing the calibration  of the $m_1$ index in terms of
[Fe/H], and $c_1$ as a function of absolute magnitude, Str\"{o}mgren
found that intermediate Population II consists of
10-15 Gyr old stars having $-0.8 <$ [Fe/H] $< -0.4$ and
velocity dispersions that are significantly
greater than those of the younger, more metal-rich disk stars.

In a seminal paper, Gilmore \& Reid (1983) showed that the distribution
of stars in the direction of the Galactic South Pole cannot be fitted
by a single exponential, but requires two disk components --
a {\it thin disk} with a scale height of 300 pc and a {\it thick disk} with
a scale height of about 1300 pc. They identified
intermediate Population II with the sum of the metal-poor end of the
thin disk and the thick disk.  Following this work, it has
been intensively discussed if the thin and thick disks
are discrete components of our Galaxy or if there is a more
continuous sequence of stellar populations connecting the
Galactic halo and the thin disk.

In another important paper, Edvardsson et al. (1993) derived precise
abundance ratios for 189 stars belonging to the disk. Stars in the temperature
range $5600 < T_{\rm eff} < 7000$\,K and somewhat evolved
away from the zero-age main-sequence were selected from
$uvby$-$\beta$ photometry of stars in the solar neighbourhood
(i.e. in  a region around the Sun with a radius of $\sim$\,100\,pc)
and divided into nine metallicity
bins ranging from [Fe/H] $\sim -1.0$ to $\sim +0.3$. In each metallicity
bin the $\sim \! 20$ brightest stars were observed. Hence, there is no kinematical
bias in the selection of the stars.

\begin{figure}
\centering
\includegraphics{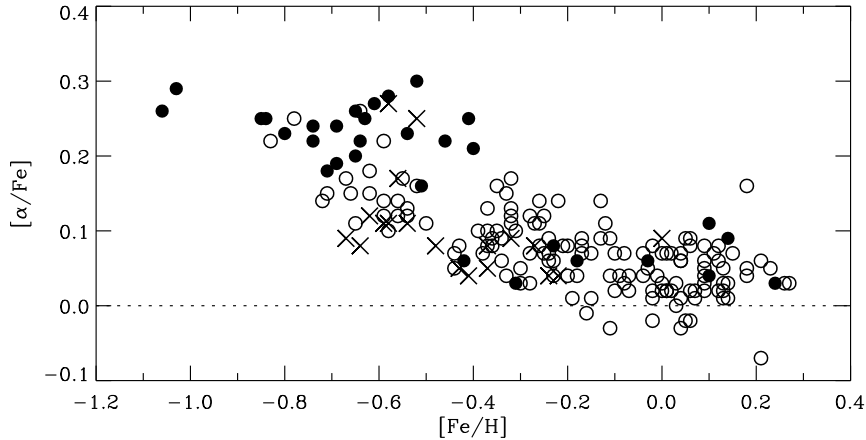}
\caption{[$\alpha$/Fe] as a function of [Fe/H] with data from
Edvardsson et al. (1993).
Stars shown with filled circles have a mean galactocentric distance in their
orbits $R_m < 7$~kpc. Open circles refer to stars with
$7 < R_m < 9$~kpc, and crosses refer to stars with $R_m > 9$~kpc.}
\label{fig:Edv.alpha-Fe.Rm}
\end{figure}

The Edvardsson et al. (1993) survey provides clear evidence for a
scatter in [$\alpha$/Fe] at a given metallicity for stars in the 
solar neighbourhood. This is
shown in Fig. \ref{fig:Edv.alpha-Fe.Rm}. As seen, [$\alpha$/Fe]
for stars in the metallicity
range $-0.8 <$ [Fe/H] $< -0.4$ is correlated with the mean
galactocentric distance $R_m$ in the stellar orbit.
Stars with $R_m > 9$~kpc tend to have lower [$\alpha$/Fe] than stars
with $R_m < 7$~kpc, and stars belonging to the solar circle lie
in between. Assuming that $R_m$  is a statistical measure of the
distance from the Galactic center at which the star was born,
Edvardsson et al. explained the [$\alpha$/Fe] variations
as due to a star formation rate that declines with
galactocentric distance. In other words, type Ia SNe
start contributing iron at a higher [Fe/H] in the inner parts
of the Galaxy than in the outer parts.

The group of stars in Edvardsson et al. (1993) with $R_m < 7$~kpc
have kinematical properties similar to those of the thick disk, for which
the dispersions of the Galactic velocity components, $U$, $V$, and $W$, 
with respect to the Local Standard of Rest (LSR)
are determined to be  ($\sigma_{U}, \sigma_{V}, \sigma_{W}) \simeq$ 
(65, 50, 40)\,km\,s$^{-1}$ ,
and for which the asymmetric drift with respect to the LSR is
$V_{\rm ad} \simeq -50$\,km\,s$^{-1}$. In comparison, thin disk stars
in the solar neighbourhood have ($\sigma_{U}, \sigma_{V}, \sigma_{W}) \simeq$
(40, 30, 20) \,km\,s$^{-1}$ and $V_{\rm ad} \simeq -10$\,km\,s$^{-1}$.
Thus, thick-disk stars move on more eccentric orbits than the thin-disk
stars, and due to the increasing density of stars towards the
inner part of the Galaxy, thick disk stars presently situated in the
solar neighbourhood tend to be close to the apo-galactic distance in their
orbits. This means that they have smaller mean galactocentric distances
than the thin-disk stars. The differences in [$\alpha$/Fe] 
shown in Fig. \ref{fig:Edv.alpha-Fe.Rm} may, therefore, also be
interpreted as due to a systematic difference in [$\alpha$/Fe]
between the thin and the thick disk. Apparently, Gratton et al. (1996)
were first to suggest this interpretation of the [$\alpha$/Fe] data.

A more clear chemical separation between thin- and thick-disk
stars has been obtained by Fuhrmann (2004).
For a sample of nearby stars with $5300 < T_{\rm eff} < 6600$\,K
and $3.7 < \log\,g < 4.6$, he derives very precise Mg abundances from Mg\,{\sc i}
lines and Fe abundances from Fe\,{\sc i} and Fe\,{\sc ii} lines.
In a [Mg/Fe] vs. [Fe/H] diagram, stars with thick disk kinematics
have [Mg/Fe] $\simeq +0.4$ and [Fe/H]
between $-1.0$ and $-0.3$. The thin disk stars show a well-defined
sequence from [Fe/H] $\simeq -0.6$ to +0.4
with [Mg/Fe] decreasing from +0.2 to 0.0. Hence, there is a
[Mg/Fe] separation between thick and thin disk stars
in the overlap region $-0.6 <$ [Fe/H] $ < -0.3$ with only a few
`transition' stars. This is even more striking in a
diagram, where [Fe/Mg] is plotted as a function of [Mg/H]
(Fuhrmann 2004, Fig.\,34).

On the basis of stellar ages derived from evolutionary tracks in
the $M_{\rm bol}$ - log\,$T_{\rm eff}$ diagram, Fuhrmann (2004) finds
that the maximum age of thin disk stars is about 9 Gyr, whereas 
thick disk stars have ages around 13\,Gyr. This suggests that
the systematic difference of [Mg/Fe] is connected to a hiatus in star formation
between the thick and thin disk phases.

\subsection{The [$\alpha$/Fe] distribution of disk stars}
\label{sect:alphadistr}

Two major studies of abundance ratios in  thin- and thick-disk stars
(Bensby et al. 2005; Reddy et al. 2003, 2006) are based
on kinematically selected groups of stars in the solar neighbourhood.
In these works, 
it is assumed that the kinematics of stars can be represented
by Gaussian distribution functions for the velocity components,
$U$, $V$, and $W$ with respect to the LSR. The kinematical probability 
that a star belongs to a given population: thin disk, thick disk or halo 
(i = 1, 2, 3) is then given by
\begin{eqnarray}
P_i \propto k_i \, f_i {\rm exp}\, (- \frac{U^2}{2 \sigma_{U_i}^2} 
- \frac{(V - V_{{\rm ad}_i})^2}{2 \sigma_{V_i}^2} - \frac{W^2}{2 \sigma_{W_i}^2}),
\end{eqnarray}
where $k = (2 \pi)^{-3/2} (\sigma_{U} \sigma_{V} \sigma_{W})^{-1}$
is the standard normalization constant,
and $f$ the relative number of stars in a given population. 
$\sigma_{U}, \sigma_{V}$, and $\sigma_{W}$  are the velocity dispersions
in $U, V$, and $W$, respectively, and $V_{\rm ad}$ the asymmetric drift velocity for
the population. As an example, the values used by Reddy et al. (2006)
are given in Table 1. Some of these values have considerable
uncertainties. Depending on how the populations is defined, the
fraction of thick disk stars in the solar neighbourhood may be as 
high as 20\,\% (Fuhrmann 2004), and the local fraction of halo stars
is often estimated to be on the order of 0.001.

\begin{table}
\centering
\caption{Velocity dispersions, asymmetric drift and fraction of stars
in the solar neighbourhood for the thin-disk, thick-disk and halo populations
as adopted by Reddy et al. (2006) for calculating membership probabilities.}
\label{table:reddy}
\setlength{\tabcolsep}{0.30cm}
\begin{tabular}{lrrrrr}
\noalign{\smallskip}
\hline\hline
\noalign{\smallskip}
Population & $\sigma_{U}$ & $\sigma_{V}$ & $\sigma_{W}$ & $V_{\rm ad}$ & $f$ \\
           &   km\,$s^{-1}$   & km\,$s^{-1}$    &  km\,$s^{-1}$   &  km\,$s^{-1}$ & \\
\hline
\noalign{\smallskip}
Thin disk     &  43  &  28   &  17  &  $-9$   & 0.93  \\
Thick disk    &  67  &  51   &  42  &  $-48$  & 0.07  \\
Halo          & 131  & 106   &  85  &  $-220$ & 0.006  \\
\noalign{\smallskip}
\hline
\end{tabular}
\end{table}

\begin{figure}
\centering
\includegraphics{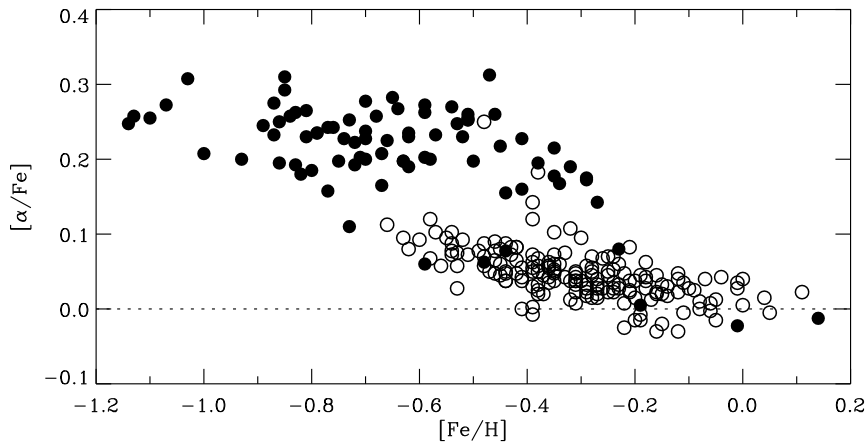}
\caption{[$\alpha$/Fe] as a function of [Fe/H] according to
Reddy et al. (2003, 2006).
Stars indicated by open circles have a probability $P > 70$\,\% of belonging to
the thin disk, whereas stars represented by filled circles have
$P > 70$\,\% of belonging to the thick disk.}
\label{fig:reddy.alpha-Fe.ThinThick}
\end{figure}

In the works of Bensby et al. (2005) and Reddy et al. (2003, 2006) precise 
abundance ratios for F and G dwarfs have been derived 
for samples of stars kinematically selected to have high probability
of belonging to either the thin or the thick disk. Thus, in Reddy et al. (2006)
the probability limit for each population is $P > 70$\,\%. The resulting
[$\alpha$/Fe] -- [Fe/H] diagram is shown in Fig. \ref{fig:reddy.alpha-Fe.ThinThick}.
As seen, there is a gap in [$\alpha$/Fe] between thin- and thick-disk stars
for the metallicity range $-0.7 <$ [Fe/H] $-0.4$. Around [Fe/H]\,$= -0.3$
a few stars seem to connect the two populations like in the corresponding
diagram of Fuhrmann (2004).

The [$\alpha$/Fe] diagram of Bensby et al. (2005) looks similar
to Fig. \ref{fig:reddy.alpha-Fe.ThinThick} except that their thick-disk
stars continue all the way up to solar metallicity with decreasing
values of [$\alpha$/Fe]. Thus, Bensby et al. claim that star formation in
the thick disk continued long enough to include chemical enrichment from
type Ia SNe. This is, however, not so evident from the data of Reddy et al.
(2006). In Fig. \ref{fig:reddy.alpha-Fe.ThinThick}, the thick disk
terminates around [Fe/H] $\simeq -0.3$ with little or no decrease of 
[$\alpha$/Fe].

Given that the thin- and thick-disk stars of Bensby et al. (2005) 
and Reddy et al. (2003, 2006) have been kinematically selected, 
the question arises
if there are stars with intermediate kinematics filling the gap in
[$\alpha$/Fe] between the two populations. The problem has be addressed
by Ram\'{\i}rez et al. (2007), who derived oxygen abundances for
523 nearby stars from a non-LTE analysis of the O\,{\sc i} triplet 
at 7774\,\AA . A similar splitting as in Fig. \ref{fig:reddy.alpha-Fe.ThinThick}
between stars with thin- and thick-disk kinematics is obtained, and
stars with intermediate kinematics do not fill the gap in [O/Fe]; they
have either high [O/Fe] or low [O/Fe].

Another set of  abundance data that can be used to study the problem
of a gap in [$\alpha$/Fe] between thin- and thick-disk stars
has been obtained by Neves et al. (2009) from ESO/HARPS
high-resolution spectra of 451 F, G, and K main-sequence stars 
in the solar neighbourhood. The main purpose of this project is to
detect planets around stars by measuring radial velocity variations
with a precision of 1\,m\,s$^{-1}$. As a by-product, stellar
abundance ratios relative to those of the Sun have been 
derived. 

\begin{figure}
\centering
\includegraphics{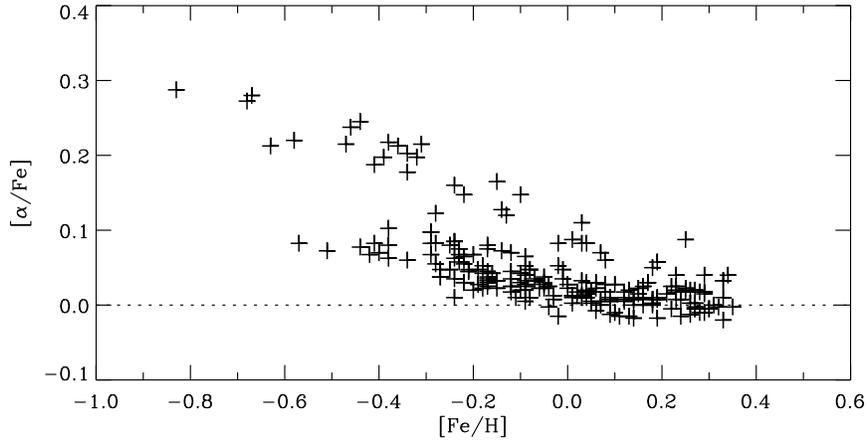}
\caption{The [$\alpha$/Fe] vs. [Fe/H] distribution for a {\em volume limited}
sample of main-sequence stars from Neves et al. (2009). Only stars with $T_{\rm eff}$
within $\pm 300$\,K from the effective temperature of the Sun have been
included.}
\label{fig:neves.alpha-Fe}
\end{figure}

Neves et al. (2009) show that trends of abundance ratios as a function
of [Fe/H] are the same for stars with and
without planets. For both groups there is a
bifurcation of the abundances of $\alpha$-capture elements
relative to iron. This is shown in Fig. 
\ref{fig:neves.alpha-Fe}, where only stars having $T_{\rm eff}$
within $\pm 300$\,K from the effective temperature of the Sun have been
included in order to obtain a very high precision of [$\alpha$/Fe],
i.e. $\sigma$\,[$\alpha$/Fe] $\simeq \pm 0.02$.
At metallicities [Fe/H] $< -0.1$, there is a gap in 
[$\alpha$/Fe] between `high-$\alpha$' and `low-$\alpha$' stars. 
Hence, the data of Neves et al.
(2009) confirm the dichotomy in [$\alpha$/Fe] found by
Bensby et al. (2005) and Reddy et al. (2006). Furthermore, the data of
Neves et al. support the claim of Bensby et al. (2005), that the thick disk
have metallicities stretching up to solar metallicity.

The sample of stars from Neves et al. (2009) is {\em volume limited};
hence, the gap between high-$\alpha$ and low-$\alpha$ stars cannot
be due to exclusion of stars with intermediate kinematics.
Figure \ref{fig:neves.alpha-Fe} shows  the distribution of the two
populations in a Toomre energy diagram.
As seen, the majority of high-$\alpha$ stars have a total space velocity
with respect to the LSR larger than 85\,km s$^{-1}$, which classify them
as thick-disk stars, but several high-$\alpha$ stars are kinematically
mixed with the low-$\alpha$ stars. This is an example of how an
abundance ratio is a more clean separator of stellar
populations than kinematics.

\begin{figure}
\centering
\includegraphics{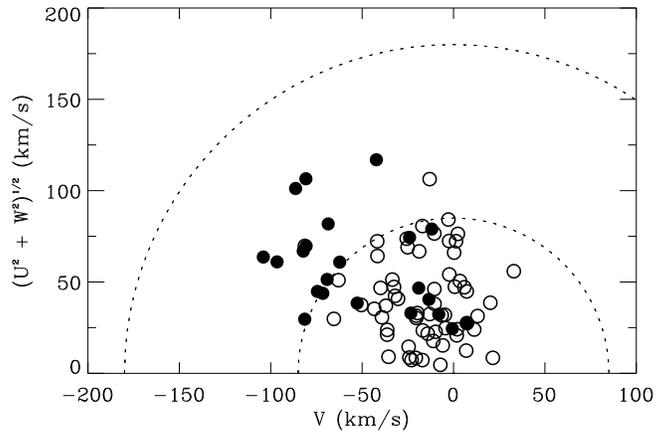}
\caption{The Toomre diagram for stars from Fig. \ref{fig:neves.alpha-Fe} with
[Fe/H] $< -0.1$. High-$\alpha$ stars 
are shown with filled circles and low-$\alpha$ stars with open circles.
The two circles delineate constant total space velocities
with respect to the LSR,
$V_{\rm tot} = 85$ and 180\,km s$^{-1}$, respectively.}
\label{fig:neves.toomre}
\end{figure}

Considering that the high-$\alpha$, thick-disk stars tend to be
older than the oldest of the low-$\alpha$, thin-disk stars (Fuhrmann 2004;
Reddy et al. 2006), the difference in [$\alpha$/Fe] is often explained
in a scenario, where a period of rapid star formation in the early 
Galactic disk was interrupted by a merging satellite galaxy that
`heated' the already formed stars to thick-disk kinematics. This was followed by
a hiatus in star formation, in which metal-poor gas was accreted and
type Ia SNe caused [$\alpha$/Fe] to decrease. When 
star formation resumed, the first thin-disk stars were
formed with low metallicity and low [$\alpha$/Fe]. This scenario also
explains the systematic differences in [C/O] (Fig. \ref{fig:CO-O.all})
and [Eu/Ba] (Fig. \ref{fig:Eu-Ba}) between thin- and thick-disk stars.

Haywood (2008) has pointed out that the low-metallicity thin-disk
stars  in the solar neighbourhood tend to have positive values of the $V$
velocity component. Such stars have mean galactocentric
distances larger than the distance of the Sun from the Galactic center, and
hence they are likely to have been formed
in the outer part of the Galactic disk.
Thick-disk stars, on the other hand, have negative values of $V$ and tend to
be formed in the inner Galactic disk. From these considerations, Haywood
suggests that the bimodal distribution in [$\alpha$/Fe] may be due to
radial mixing of stars in the disk.

This scenario has been further investigated by Sch\"{o}nrich \& Binney 
(2009a, 2009b) in a model for the chemical evolution of the Galactic
disk, for which the star formation rate is monotonically decreasing, and
which includes radial migration
of stars and gas flows. The model successfully fit the 
metallicity distribution and the large scatter in the
age-metallicity relation for stars in the 
Geneva-Copenhagen survey (Nordstr\"{o}m et al. 2004). The model also
predicts a bimodal distribution of [$\alpha$/Fe] for stars in the solar
neighbourhood with the high-$\alpha$ stars coming from the inner parts of
the Galactic disk and the low-$\alpha$ stars from the outer part. It will,
however,
be interesting to see if the model can reproduce a gap in the [$\alpha$/Fe]
distribution for disk stars as found from the data of Neves et al. (2009).

\subsection{Abundance gradients in the disk}
\label{sect:gradients}

In models for the chemical evolution of the Galactic disk, observed
abundance gradients provide important constraints.
Gradients may be determined from 
B-type stars and H\,II regions, but the most precise results have
been obtained from Cepheids. These variable stars are bright
enough to be studied at large distances and accurate values of the
distances can be obtained from the period-luminosity relation. 
Earlier results suggested a steeper metallicity gradient in
the inner part of the galaxy  as compared to the outer part with a
break in the gradient occurring around 10\,kpc. According to newer
work this is, however, not so obvious. For 54 Cepheids ranging
in galactocentric distance $R_G$ from 4 to about 14\,kpc,  Luck et al. (2006)  
find an overall gradient $d$[Fe/H]/$dR_G = -0.06$\,dex kpc$^{-1}$,
but there seems to be a significant cosmic scatter around a linear
fit to the data. A region at Galactic longitude $l \sim 120$ degrees
and a distance of 3-4\,kpc from the Sun has enhanced
metallicities with $\Delta$\,[Fe/H]\,$\simeq +0.2$ in the mean.
Such spatial inhomogeneities could be due to recent SNe events.

Yong et al. (2006) have made an interesting study of 24 Cepheids in
the outer Galactic disk based on high-resolution spectra. The 
distance range is 12\,$< \, R_G \,< 18$\,kpc. Most of the Cepheids
continue the trend with galactocentric distance exhibited by the
Luck et al. (2006) sample, but a minority of six Cepheids
have [Fe/H] around $-0.8$\,dex and enhanced $\alpha$-element
abundances, [$\alpha$/Fe]\,$\sim +0.3$. Thus, there is some
evidence for two populations of Cepheids in the outer disk.

Cepheids can only provide information about `present-day' 
gradients in the disk. Abundances of stars in open clusters may,
however, be used to determine gradients at different ages. 
In this connection one benefits from the quite accurate 
ages and distances that can be determined from colour-magnitude
diagrams of open clusters. For a review of results from clusters,
the reader is referred to the chapter on open clusters
by Friel (this volume).

\section{The Galactic bulge}
\label{sect:bulge}
Due to a large distance and a high degree of interstellar absorption
and reddening, the Galactic bulge is the least well known component
of the Milky Way. It has been much discussed if the bulge 
contains only very old stars or if it also includes a younger population.
This problem is related to two different scenarios for the formation of
the bulge,  a `classical' bulge formed rapidly 
by the coalescence of star-forming clumps as suggested from the simulations of
Elmegreen  et al. (2008) or a  `pseudobulge' formed over a 
longer time through dynamical instabilities in the Galactic disk
(see review by Kormendy \& Kennicutt 2004). The measurement of
abundance ratios in bulge stars may help to decide between these
scenarios by providing information on the time-scale for the formation
of the bulge.

The metallicity distribution of bulge stars has been debated for a long
time. With the determination of [Fe/H] for 800 bulge giants 
based on VLT multi-fiber spectra with a resolution of $R \sim 20\,000$ 
(Zoccali et al. 2008) a robust result seems to have been obtained.
Zoccali et al. observed stars in four fields having distances from
the Galactic center ranging from 600 to 1600\,pc. The overall 
metallicity distribution is centered on solar metallicity and
extends from [Fe/H]\,$\simeq -1.5$ to +0.5, but with few stars in
the range $-1.5 <$ [Fe/H] $< -1.0$. A decrease of the mean metallicity
along the bulge minor axis is suggested corresponding to a gradient
of $\sim \! 0.6$\,dex per kpc. 

A pioneering study of $\alpha$/Fe ratios in 12 bulge red giants was
carried out by McWilliam \& Rich (1994) based on 4-meter telescope optical
echelle spectra with a resolution of $R \sim 20\,000$ and typical
$S/N \sim 50$. Enhanced values of [Mg/Fe] and [Ti/Fe] ($\sim \! +0.3$\,dex)
were found for all stars up to a metallicity of $\sim \! +0.4$\,dex
suggesting a very rapid formation of the bulge. In contrast, [Si/Fe] and [Ca/Fe] 
showed solar values, which is difficult to understand. These data are
now superseeded by higher resolution spectra obtained with 8-meter 
class telescopes both in the optical and the infrared spectral region.

Lecureur et al. (2007) used VLT/UVES spectra
in the spectral region 4800 - 6800\,\AA\ to derive O and Mg abundances
relative to Fe for about 50 red bulge giants. They find that the [O/Fe] and
[Mg/Fe] trends are more enhanced than the corresponding trends for
thick-disk stars as determined by Bensby et al. (2005) for
dwarf stars. The same conclusion is reached by Fulbright et al. (2007)
from abundance ratios derived for 27 red giants observed at
high resolution in the 5000 - 8000\,\AA\ spectral region with the Keck/HIRES
spectrograph.

\begin{figure}
\centering
\includegraphics{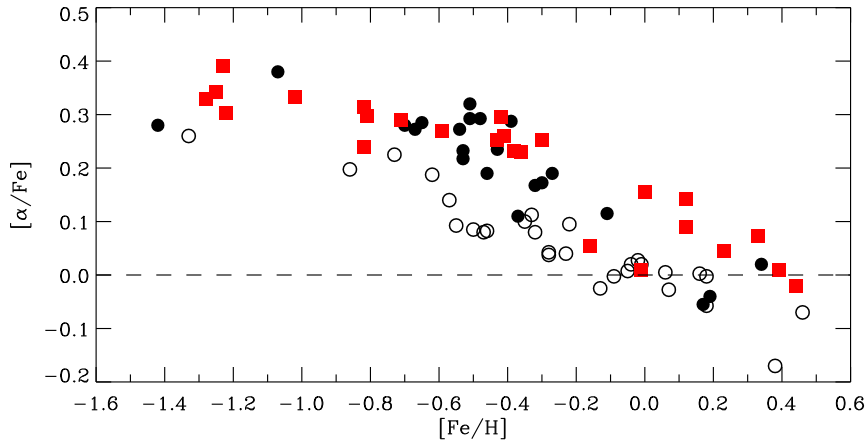}
\caption{The [$\alpha$/Fe] - [Fe/H] relation for bulge K giant stars
shown with filled (red) squares in comparison with K giants in the
solar neighbourhood having either thick-disk kinematics (filled circles)
or thin-disk kinematics (open circles). Data for [Mg/Fe], [Si/Fe],
[Ca/Fe], and [Ti/Fe] are adopted from Alves-Brito et al. (2010).}
\label{fig:bulge.alpha-fe}
\end{figure}

In contrast to these results, Mel\'{e}ndez et al. (2008) find the same
[O/Fe] - [Fe/H] trend for 19 red giants belonging to the bulge and 
21 thick-disk giants in the solar neighbourhood. For both samples,
the oxygen and iron abundances are based on spectrum synthesis of
OH and Fe\,{\sc i} lines in an infrared spectral region around 
1.55\,$\mu$m as observed with Gemini/Phoenix. Hence, the comparison of
the bulge and the thick disk is done in a differential way for the same
type of stars. Mel\'{e}ndez et al. suggest that the comparisons
of bulge giants with thick-disk dwarfs has led to
a spurious offset of the two [O/Fe] - [Fe/H] trends due to systematic errors
in the abundance ratios. When LTE
is assumed and classical 1D model atmospheres are adopted,
systematic errors could arise due to different non-LTE and 3D 
corrections for dwarf and giant stars. 
Recently, Ryde et al. (2010) have confirmed the results
of Mel\'{e}ndez et al. (2008) from VLT/CRIRES high-resolution spectra of
11 red giants in the bulge.

Further evidence for an agreement between bulge and local thick-disk stars
has been obtained by Alves-Brito et al. (2010) from
optical high-resolution spectra.
Their results are shown if Fig. \ref{fig:bulge.alpha-fe}.
As seen the trends for bulge and 
thick-disk stars agree well with no significant differences below
solar metallicity. The thin-disk stars, on the other hand, 
fall below the bulge and the thick disk stars. Above solar
metallicity, the bulge stars tend to have higher [$\alpha$/Fe]
values than the disk stars but
this needs to be confirmed for a larger sample.

Independent and very interesting data for abundances
in the Galactic bulge have been obtained from
spectra of microlensed main-sequence
and subgiant stars. During a microlensing event, the flux from a
star can be enhanced by a factor of 100 or even more;  hence, the magnitude of
a typical bulge turnoff star raises from $V \sim 18$ to $V \sim 13$. This
makes it possible to obtain high-resolution spectra with good $S/N$.
Bensby et al. (2011) have presented a homogeneous abundance study of 26
such microlensed stars using the same methods as applied for local
thick-disk dwarf stars (Bensby et al. 2005). As seen from
Fig. \ref{fig:all.alpha-fe}, there is a tendency that these
microlensed bulge stars are separated into two regions in the 
[$\alpha$/Fe] -- [Fe/H] plane: A metal-poor group with
[Fe/H] $< -0.3$ and [$\alpha$/Fe] $\simeq +0.3$ like the
thick-disk stars, and a metal-rich group with +0.1 $<$ [Fe/H] $<$ +0.6
for which the majority of stars have [$\alpha$/Fe] $\simeq 0$ like the
thin-disk stars, but four stars have [$\alpha$/Fe] $\simeq +0.1$. 

Another interesting aspect of the study of microlensed main-sequence
and subgiant stars is the possibility to derive ages by comparing their
position in a log$g$ - log$T_{\rm eff}$ diagram with isochrones computed
from stellar models.  According to Bensby et al. (2011) the metal-poor
bulge stars have an average age of 11.2\,Gyr with a dispersion 
of $\pm 2.9$\,Gyr much of which may be due errors in the age determination.
The metal-rich bulge stars are on average younger (7.6\,Gyr) and
has a larger age dispersion ($\pm 3.9$\,Gyr).  

On the basis of these new data, Bensby et al. (2011) suggest that
the bulge consists of two stellar populations, i.e. a metal-poor population
similar to the thick disk in terms of metallicity range, 
[$\alpha$/Fe], and age, and a metal-rich population, which could
be related to the inner thin disk. Supporting evidence has recently
been obtained by Hill et al. (2011) from a study of 219 red giants
in Baade's bulge window (situated about four degrees from the
Galactic center). [Fe/H] and [Mg/Fe] are derived from ESO VLT
FLAMES spectra with a resoltion of $R =$ 20\,000. The distribution
of [Fe/H] is asymmetric and can be deconvolved into two Gaussian
components: a metal-poor population centred at [Fe/H] = $-0.30$
having a dispersion of 0.25\,dex in [Fe/H],
and a metal-rich population centred at
[Fe/H] = $+0.32$ with a dispersion of 0.11\,dex only.
The metal-poor population has [Mg/Fe] $\simeq 0.3$, whereas
the stars in the metal-rich population distribute around the
solar Mg/Fe ratio. These data agree well with those of 
Bensby et al. (2011).
However, a larger set of data for microlensed
bulge stars and information about chemical abundances of inner
disk stars are needed before one can draw any robust
conclusions about the existence of two  bulge populations and the
consequences this may have for models of the formation of the bulge. 

\section{The Galactic halo}
\label{sect:halo}

\subsection{Evidence of two distinct halo populations}
\label{sect:twohalopop}

For a long time, it has been discussed
if the Galactic halo consists of more than one population. The classical
monolithic collapse model of Eggen, Lynden-Bell \& Sandage (1962)
corresponds to a single halo
population, but from a study of globular clusters, Searle and Zinn
(1978) suggested that the halo comprises
two populations: $i)$ an inner, old, flattened
population with a slight prograde rotation formed during a dissipative
collapse, and $ii)$ an outer, younger, spherical population
accreted from satellite systems.
This dichotomy of the Galactic halo has found support in
a study of $\sim\,$20\,000 stars from the Sloan Digital Sky Survey (SDSS)
by Carollo et al. (2007).  They find that the inner halo consists of stars with
a peak metallicity at [Fe/H]\,$\simeq -1.6$, whereas the outer halo
stars distribute around [Fe/H]\,$\simeq -2.2$. 

Several studies suggest that there is a difference in
[$\alpha$/Fe] between stars that can be associated with the inner 
and the outer halo, respectively.
Fulbright (2002) shows that stars with large
values of the total space velocity relative to the LSR,
$V_{\rm tot} > 300$\,km\,s$^{-1}$, tend to have lower values of [$\alpha$/Fe] than
stars with $150 < V_{\rm tot} < 300$\,km\,s$^{-1}$, and Stephens \& Boesgaard
(2002) find that [$\alpha$/Fe] is correlated with the apo-galactic
orbital distance in the sense that the outermost 
stars have the lowest values of [$\alpha$/Fe]. Further support
for such differences in [$\alpha$/Fe] comes from a study 
by Gratton et al. (2003), who divided stars in the solar neighbourhood 
into two populations according to their
kinematics: $i)$ a `dissipative' component, which includes 
thick-disk stars and prograde rotating halo stars, and $ii)$ an
`accretion' component that consists of retrograde rotating halo
stars. The `accretion' component has smaller values and a larger
scatter for [$\alpha$/Fe] than the `dissipative' component. 

The differences in [$\alpha$/Fe] found in these works are
not larger than  about 0.1\,dex, and it is unclear, if the
distribution of [$\alpha$/Fe] is continuous or bimodal. 
Nissen \& Schuster (1997) found a more clear dichotomy in [$\alpha$/Fe]
for 13 halo stars with  $-1.3 <$ [Fe/H] $< -0.5$. Eight of these halo
stars have [$\alpha$/Fe] in the range 0.1 to 0.2\,dex,
whereas [$\alpha$/Fe] $\simeq 0.3$ for the other five halo stars.
Interestingly, the low-$\alpha$ halo stars
tend to have larger apo-galactic distances than the high-$\alpha$ stars.

A more extensive study of `metal-rich' halo stars has been
carried out by Nissen \& Schuster (2010). Stars are selected from the 
Schuster et al. (2006) $uvby$-$\beta$ catalogue of high-velocity and metal-poor
stars. To ensure that a star has a high probability of belonging to the halo
population, the total space velocity with respect to
the LSR is required to be larger than 180\,km\,s$^{-1}$.
Furthermore, Str\"{o}mgren photometric indices
are used to select dwarfs and subgiants with
$5200 < T_{\rm eff} < 6300$\,K and [Fe/H] $> -1.6$.
High-resolution, high $S/N$ spectra were obtained with the 
ESO VLT/UVES and the Nordic Optical Telescope FIES spectrographs for 78 of these stars.
The large majority of the stars are brighter than $V = 11.1$ and situated
within a distance of 250\,pc. These spectra are used to derive
high-precision LTE abundance ratios in a differential analysis that
also includes 16 stars with thick-disk kinematics. The precision of
the various abundance ratios range from 0.02 to 0.04\,dex.

\begin{figure}
\centering
\includegraphics{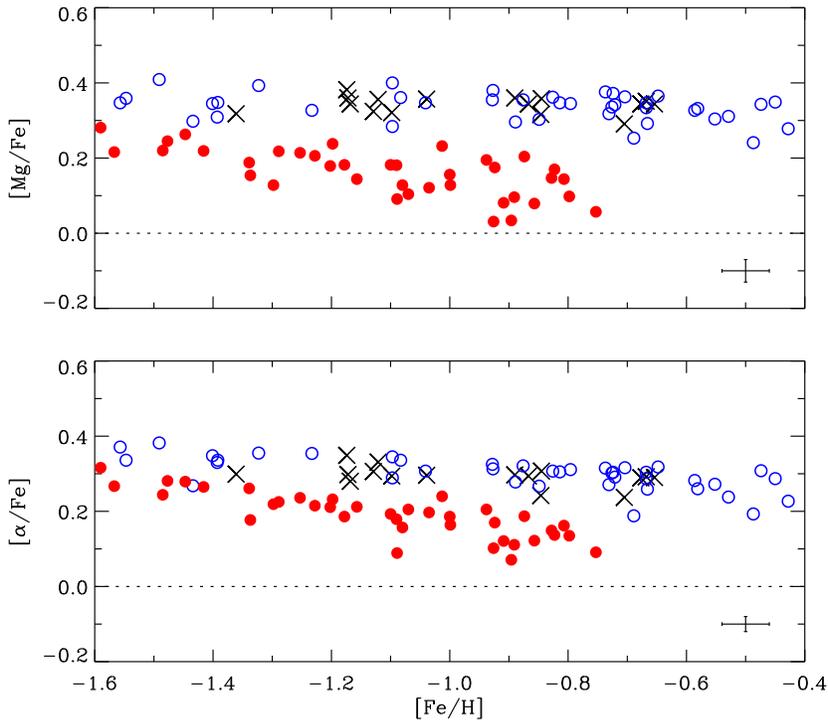}
\caption{[Mg/Fe] and [$\alpha$/Fe] versus [Fe/H] based on
data from Nissen \& Schuster (2010). Stars with
thick-disk kinematics are indicated by crosses and halo stars
by circles. On the basis of the [Mg/Fe] distribution, 
the halo stars are divided into low-$\alpha$
stars shown by filled (red) circles, and high-$\alpha$ stars shown
by open (blue) circles.}
\label{fig:nissen.mg.alpha-fe}
\end{figure}

Figure \ref{fig:nissen.mg.alpha-fe} (top) shows the distribution of
[Mg/Fe]  as a function of [Fe/H] for stars in Nissen \& Schuster (2010).
As seen, the halo stars split into two distinct populations:
`high-$\alpha$' stars with a nearly constant [Mg/Fe] and
`low-$\alpha$' stars with declining values of [Mg/Fe] 
as a function of increasing metallicity. A classification into these 
two populations has been done on the basis of [Mg/Fe], but as seen
from the bottom part of the figure, [$\alpha$/Fe] would have led to
the same division of the halo stars except at the lowest metallicities,
$-1.6 <$ [Fe/H] $< -1.4$, where the two populations tend to merge, 
and the classification is less clear.

As seen from Fig. \ref{fig:nissen.mg.alpha-fe},
the separation in [Mg/Fe] for the two halo populations is significantly
larger than the separation in [$\alpha$/Fe]. 
At [Fe/H]\,$\simeq  -0.8$ the mean difference in [Mg/Fe] is about 0.25\,dex,
whereas it is only about 0.15\,dex in [$\alpha$/Fe]. 
This is probably caused by different degrees of SNe Ia
contribution to the production of Mg, Si, Ca, and Ti. According to
Tsujimoto et al. (1995, Table 3),
the relative contribution of SNe Ia to the solar composition 
is negligible for Mg, 17\% for Si and 25\% for Ca (Ti was not included).
For comparison, the SNe Ia contribution is 57\% for Fe.
Hence, [Mg/Fe] is a more sensitive measure of the ratio between type II  
and type Ia contributions to the chemical enrichment of matter than 
[$\alpha$/Fe]. On the other hand, it is possible to measure [$\alpha$/Fe] 
with a higher precision than [Mg/Fe], because many more spectral lines
can be used to determine [$\alpha$/Fe].  

\begin{figure}
\centering
\includegraphics{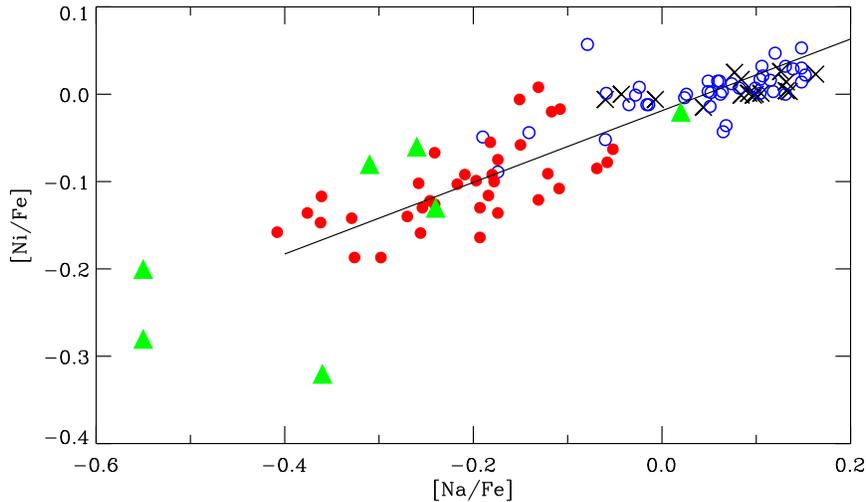}
\caption{The relation between [Ni/Fe] and [Na/Fe]. Local
main-sequence stars from Nissen \& Schuster (2010) are indicated by the same 
symbols as in Fig. \ref{fig:nissen.mg.alpha-fe}, whereas the (green) triangles
refer to K giants in dSph satellite galaxies from
Venn et al. (2004). For both sets of data,
the stars are confined to the metallicity range $-1.6 <$ [Fe/H] $< -0.4$.}
\label{fig:ni-na.all}
\end{figure}

The low-$\alpha$ halo stars also have low values of [Na/Fe] and [Ni/Fe]
relative to the high-$\alpha$ stars, and as shown in Fig. 
\ref{fig:ni-na.all}, the two populations are well separated
in a [Ni/Fe] - [Na/Fe] diagram. As seen, some stars in dSph galaxies 
are even more extreme in [Na/Fe] and [Ni/Fe] than the low-$\alpha$ halo stars.

\subsection{Kinematics and origin of the two halo populations}
\label{sect:kinorig}

The distribution of [$\alpha$/Fe] in  Fig. \ref{fig:nissen.mg.alpha-fe}
can be explained, if the high-$\alpha$ stars have been formed in regions
with such a high rate of chemical evolution that only
type II SNe have contributed to the chemical enrichment up to 
[Fe/H]\,$\sim -0.4$. 
The low-$\alpha$ stars, on the other hand, originate from
regions with a relatively slow chemical evolution so that
type Ia SNe have started to contribute iron around
[Fe/H]\,=\,$-1.6$ causing [$\alpha$/Fe] to decrease towards higher
metallicities until [Fe/H]\,$\sim -0.8$.

\begin{figure}
\centering
\includegraphics{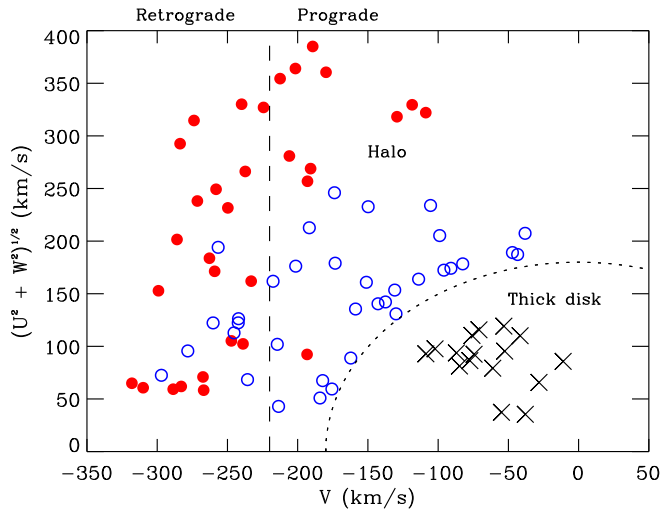}
\caption{Toomre diagram for stars from Nissen \& Schuster (2010)
having [Fe/H]\,$> -1.4$.
The same symbols for high-$\alpha$ halo, low-$\alpha$ halo and thick-disk stars
as in Fig. \ref{fig:nissen.mg.alpha-fe} are used. The short-dashed circle corresponds
to $V_{\rm tot} = 180$\,km s$^{-1}$. The long-dashed line indicates 
zero rotation in the Galaxy and therefore separates retrograde moving
stars from prograde moving.}
\label{fig:toomre.review}
\end{figure}

Further insight into the origin of the two halo populations can
be obtained from kinematics.
As seen from the Toomre energy diagram in  Fig. \ref{fig:toomre.review},
the high-$\alpha$ stars show
evidence for being more bound to the Galaxy and
favoring prograde Galactic orbits, while the low-$\alpha$
stars are less bound with two-thirds of them being on
retrograde orbits. This suggests that the high-$\alpha$
population is connected to a dissipative component of the Galaxy,
while the low-$\alpha$ stars have been accreted from dwarf galaxies.

\begin{figure}
\centering
\includegraphics{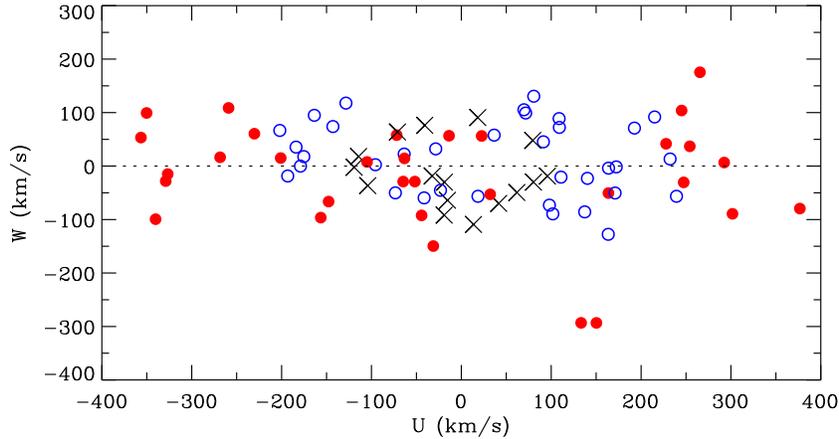}
\caption{The relation between the $U$ and $W$ velocity components
for stars from Nissen \& Schuster (2010) having [Fe/H]\,$> -1.4$.}
\label{fig:W-U.review}
\end{figure}

Several retrograde moving low-$\alpha$ stars
have a Galactic $V$-velocity component similar to that of
the $\omega$\,Cen globular cluster, i.e. $V \sim -260$\,km s$^{-1}$.
As often discussed (e.g. Bekki \& Freeman 2003), $\omega$\,Cen is
probably the nucleus of a captured satellite galaxy with its
own chemical enrichment history. Meza et al. (2005)
have simulated the orbital characteristics of the tidal debris of
such a satellite dragged into the
Galactic plane by dynamical friction. The captured stars
are predicted to have rather small $W$-velocities but a wide, double-peaked
$U$-distribution. As shown in Fig. \ref{fig:W-U.review}, the $W$-$U$  distribution
observed for the low-$\alpha$ halo  corresponds quite well to that
prediction. There are two groups of low-$\alpha$ stars with $U > 200$\,km s$^{-1}$
and $U < -200$\,km s$^{-1}$, respectively, corresponding to stars
moving in and out of the solar neighbourhood on elongated radial orbits.
Thus, a good fraction of the low-$\alpha$ stars, although not all, may
well have originated in the $\omega$\,Cen progenitor galaxy.
The high-$\alpha$ stars, on the other hand, are confined to a
much smaller range in $U$, i.e. from about $-200$ to about +200\,km s$^{-1}$.

The [$\alpha$/Fe] vs. [Fe/H] trend for the low-$\alpha$ stars  
in Fig. \ref{fig:nissen.mg.alpha-fe} and the 
Ni-Na trend in  Fig. \ref{fig:ni-na.all} resemble the 
corresponding trends for stars in dwarf galaxies.
Stars in these systems tend, however, to have lower values of
[$\alpha$/Fe], [Na/Fe] and [Ni/Fe] than
low-$\alpha$ halo stars. This offset agrees with simulations of the
chemical evolution of a hierarchically formed stellar halo in a
$\Lambda$CDM Universe by Font et al. (2006, Fig. 9).
The bulk of halo stars originate from early accreted,
massive dwarf galaxies with efficient star formation, whereas
surviving satellite galaxies in the outer halo on average have
smaller masses and a slower chemical evolution with a larger
contribution from Type Ia SNe at a given metallicity.
The [Mg/Fe] vs. [Fe/H] trend for field stars, predicted by
Font et al., agrees in fact remarkably
well with the trend for the low-$\alpha$ halo stars. 

The simulations of Font et al (2006) do not explain the existence 
of high-$\alpha$ halo stars.
Two recent $\Lambda$CDM simulations suggest, however, a dual origin
of stars in the inner Galactic halo. Purcell et al. (2010)
propose that ancient stars formed in the Galactic disk can be ejected
to the halo by merging satellite galaxies, and
Zolotov et al. (2009, 2010) find that stars formed out of accreted
gas in the inner 1\,kpc of the Galaxy can be displaced into the halo through
a succession of mergers. Alternatively, the high-$\alpha$ population
may simply belong to the high-velocity
tail of a thick disk with a non-Gaussian velocity distribution.

\subsection{Globular clusters and dwarf galaxies}
\label{sect:globclusters}
The Galactic halo contains more than 100 globular clusters and a 
number of dwarf spheroidal galaxies that are considered as 
Milky Way satellite systems. A dSph galaxy has a broad range 
in age and metallicity, whereas a globular cluster is a smaller system
with a more limited range in these parameters . As often discussed,
it is possible  that  some or all field halo stars come from
dissolved globular clusters or have been accreted from previous
generations of satellite galaxies. In this connection, it is is
of great interest to compare the chemical abundance ratios in 
field stars with the corresponding ratios in still existing
globular clusters and dwarf galaxies.

\begin{figure}
\centering
\includegraphics{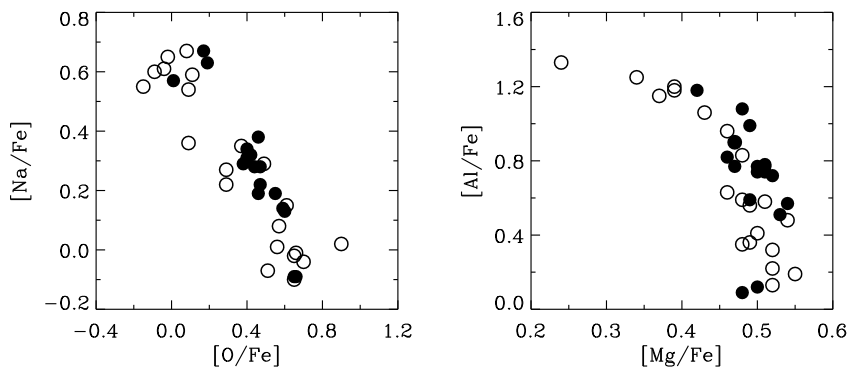}
\caption{The Na-O and Al-Mg anti-correlations in the globular
cluster NGC\,6752 based on data from Yong et al. (2005). Filled
circles indicate stars near the bright end of the red-giant branch;
open circles refer to less luminous stars around the red-giant bump.}
\label{fig:NGC6752}
\end{figure}

Globular cluster were for a long time considered as examples of
systems containing a single population of stars with a well defined
age and chemical composition. During the last decades it has,
however, become more and more clear that many, if not all, globular
clusters contain multiple stellar populations. Strong evidence
comes from abundance ratios between elements
from oxygen to aluminum. As reviewed by Gratton et al. (2004),
high-resolution spectroscopy of red giants has revealed anti-correlations
between [Na/Fe] and [O/Fe] and between [Al/Fe] and [Mg/Fe] 
in intermediate metallicity globular clusters. An example is shown in 
Fig. \ref{fig:NGC6752} for K giants in  NGC\,6752 (Yong et al. 2005).

The extensive study of Carretta et al. (2009) based on high-resolution
VLT/UVES spectra for giant stars in 19 globular clusters indicates
the existence of a [Na/Fe] - [O/Fe] anti-correlation in all cases,
but the amplitude of the variations is different from cluster to
cluster. Variations in [Al/Fe] correlated with [Na/Fe] 
are seen in the majority of clusters and an anti-correlation
between [Al/Fe] and [Mg/Fe] is detected in a few cases.
In addition, Yong et al. (2003) 
have revealed a correlation between the heavy magnesium isotope 
$^{26}$Mg and Al in NGC\,6752 by determining Mg isotope ratios 
from the profiles of MgH lines near 5140\,\AA .

All of these abundance variations are ascribed to hydrogen burning
via the CNO-cycle and the NeNa and MgAl chains. Some years ago,
it was much discussed if the variations were due to nuclear processes and
non-standard mixing in the stars themselves or to an early generation 
of stars that have polluted the gas out of which the present low-mass
stars in the clusters have formed. The fact that the abundance variations
are not correlated with the luminosity of the giant stars speaks against
the  first possibility, and when Gratton et al. (2001) discovered that
a Na-O anti-correlation is present in turnoff and subgiant stars in
NGC\,6397 and NGC\,6752, the internal mixing case was ruled out. Hence,
the abundance variations must be due to an early generation of stars.
Candidates are intermediate-mass AGB stars undergoing hot bottom
burning (Ventura et al. 2001) and massive rotating stars (Decressin et al.
2007). 

In addition to the abundance variations of elements from O to Al,
there is also increasing evidence for variations of Ca, Fe and $s$-process
elements in several globular clusters. The classical example is $\omega$\,Cen
for which giant stars are found to have metallicities from [Fe/H]\,$\simeq -1.9$
to $\simeq -0.5$ (see review by Gratton et al. 2004). Multiple sequences
in colour-magnitude diagrams are present suggesting the existence of
four or five discrete populations in $\omega$\,Cen with significant age and
abundance differences.  
The more metal-rich stars in $\omega$\,Cen have unusually
low values of [Cu/Fe] (Cunha et al. 2002) and very high values of
the abundances of the second-peak $s$-process elements, i.e. 
[Ba/Fe] and [La/Fe] values around 1.0\,dex (Smith et al. 2000).
These abundance 
anomalies point to a complicated chemical evolution history, in which
$\omega$\,Cen was originally the nucleus of a much larger dwarf galaxy 
that merged with the early Galactic disk (Bekki \& Freeman 2003).

Evidence for abundance variations of Ca, Fe, and  
$s$-process elements has been found for other clusters than $\omega$\,Cen. 
On the basis of high-resolution spectroscopy with UVES of eight 
giant stars in NGC\,1851, Yong \& Grundahl (2008) found a range of 
0.6\,dex in [Zr/Fe] and [La/Fe], and
Marino et al. (2009) found similar large variations in [Y/Fe],
[Zr/Fe] and [Ba/Fe] for 17 giant stars in M\,22. In addition,
there is a correlation between these ratios and [Fe/H], which 
varies from $-1.9$ to $-1.5$ in M\,22.
Evidence of variations in [Ca/Fe] is also found
in other globular clusters by Lee et al. (2009) from 
photometric measurements of an index of the Ca\,{\sc ii} H and K
lines, but Carretta et al. (2010) do not confirm this on the basis of
high-resolution UVES spectra. Instead, they suggest that the spread
in the Ca\,{\sc ii} H and K index may be related to variations in
helium and nitrogen abundances.

Some authors, e.g. Lee et al. (2009), suggest that all globular 
clusters were once nuclei of dwarf galaxies that are now accreted 
and dissolved
in the Milky Way. In this connection, one may wonder why practically
no field halo stars share the Na-O abundance anomalies of the
globular clusters. Perhaps the explanation is that the 
elements produced by AGB stars are confined to the potential
wells of the clusters. According to the hydrodynamical
simulations by D'Ercole et al. (2008), the gas ejected by
AGB stars collects in cooling flows into the cores of globular
clusters.

\begin{figure}
\centering
\includegraphics{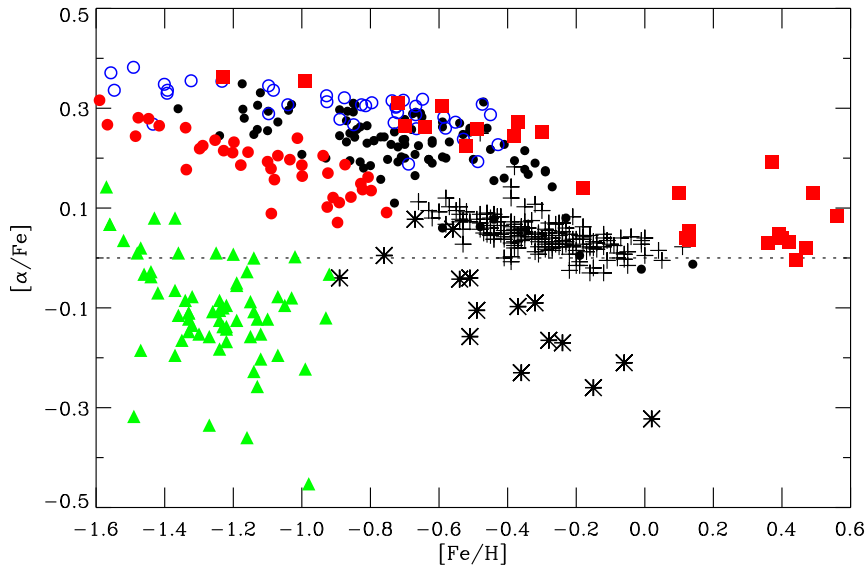}
\caption{[$\alpha$/Fe] vs. [Fe/H] for various stellar populations.
Thin-disk stars from Reddy et al. (2003) are shown with plus symbols.
Filled circles refer to thick-disk stars from Reddy et al. 
(2006) and Nissen \& Schuster (2010).  
Filled (red) squares are microlensed bulge stars from
Bensby et al. (2011). 
Open (blue) circles are high-$\alpha$ and filled (red) circles
low-$\alpha$ halo stars from Nissen \& Schuster (2010).
Asterisks refer to stars in the Sagittarius dSph galaxy (Sbordone et al. 2007),
and filled (green) triangles show data for stars in the Sculptor dSph galaxy
from Kirby et al. (2009) for which the precision of [$\alpha$/Fe] is
better than 0.15\,dex.}
\label{fig:all.alpha-fe}
\end{figure}

The chemical composition of giant stars in dSph galaxies is
reviewed by Tolstoy et al. (2009). For the large majority of stars
with [Fe/H]\,$> -2$, [$\alpha$/Fe] is significantly lower than in
halo and disk stars belonging to the Milky Way. Two examples are shown
in Fig. \ref{fig:all.alpha-fe}. Stars in Sculptor (Kirby et al. 2009)
distribute around [$\alpha$/Fe] = $-0.1$, and do not seem to have
metallicities above [Fe/H]\,$\sim -0.9$. The other example is the
Sagittarius dwarf galaxy, which was discovered by Ibata et al. (1994)
to be merging with the Milky Way. The abundances derived by
Sbordone et al. (2007) for giant stars show 
a declining trend of [$\alpha$/Fe] 
reaching [$\alpha$/Fe] as low as $-0.3$\,dex at
solar metallicity. 

Evidently, present day dSph galaxies have had a chemical evolution
history different from that of the other main components of
the Milky Way. In addition to the underabundance of [$\alpha$/Fe], the 
more metal-rich stars in dSph galaxies stand out
by having low [Na/Fe] and [Ni/Fe] abundances (Fig. 11) as well as 
high [Ba/Y] ratios (Venn et al. 2004). At the lowest metallicities,
[Fe/H]\,$< -2$, the abundance ratios are, however, similar to those
in Galactic halo stars, which suggests that the abundance deviations
of dSph stars should not be explained in terms of an anomalous 
IMF. The most obvious reason for the
underabundances of [$\alpha$/Fe] is that the star formation rate
in surviving dSph galaxies has been so slow that type Ia SNe started to
contribute Fe at a metallicity of [Fe/H]\,$\sim -2$. Dwarf galaxies
are, however, not ruled out as `building blocks' of the Galactic halo,
because early accreted dSph galaxies probably have had a somewhat faster 
chemical evolution than the less masive present-day dSph galaxies
in the outer halo (Font et al. 2006).

\section{Conclusions}
\label{sect:conclusions}
The last decade has seen great progress in determinations of
abundance ratios that can be used as tracers of
stellar populations. This has been possible due to the availability of 
high-resolution echelle spectrographs at large telescopes -- in some
cases with multiplex capabilities so that many stars, e.g. in clusters
and dwarf galaxies, can be observed simultaneously. Determinations 
of [$\alpha$/Fe] from low- or medium-resolution spectroscopy,
e.g. in the SDSS and RAVE surveys, are also important
in connection with large statistical investigations of stellar populations.
Similar measurements of [$\alpha$/Fe] will be obtained in 
connection with the GAIA mission. Furthermore,
abundances of Galactic bulge stars are now been obtained from 
high-resolution infrared spectra, but the spectral
coverage is small and there is no multiplex advantages. Hence,
the number of bulge stars studied with infrared spectra is still
limited. More efficient infrared echelle spectrographs with multiplex 
capabilities are needed to take advantage of the lower reddening in 
the infrared spectral region, when dealing with the bulge and the
inner part of the disk.

Elemental abundance ratios are derived by the aid of a 
model atmosphere analysis of the available stellar spectra as discussed in 
Sect. \ref{sect:abundet}. Most studies are still based on homogeneous
model atmospheres and the assumption of LTE, but several works show that
inhomogeneous models and deviations from LTE can change the derived abundances
significantly. Such 3D and non-LTE modelling is difficult to carry out
and in some cases the derived abundances are sensitive
to uncertain hydrogen collision cross sections. Hence, the derived
trends of abundance
ratios as a function of metallicity can be quite uncertain. 
On the other hand, it is possible to determine precise differences
of abundance ratios by a 1D LTE analysis of a sample of stars
confined to small ranges in effective temperature, surface gravity and metallicity.
In this way, abundance ratios can be used to disentangle stellar
populations. Care should, however, be taken when comparing different
spectral types and luminosity classes, such as F and G dwarfs with
K giants, because systematic errors may be important.

As discussed in Sect. \ref{sect:tracers}, the abundance ratios
[$\alpha$/Fe], [C/O], [Na/Fe], [Ni/Fe], [Cu/Fe], [Eu/Ba], and [Ba/Y]
are of high interest as tracers of stellar populations.
The usefulness of these ratios is related to the fact
that the elements involved are produced in different types of stars. The 
nucleosynthesis is not well understood for all elements, but it
seems that variations of the ratios from one population to the
next can be explained in terms of different rates of star formation
and chemical evolution. In general, one does
not need to invoke variations in the initial mass function to explain
the abundance ratios. A better understanding of the nucleosynthesis
of the elements is, however, important in order to learn more about
the formation and evolution of the various populations.

The power of abundance ratios as population tracers is evident from
Fig. \ref{fig:all.alpha-fe}, where [$\alpha$/Fe] is plotted as
a function of [Fe/H] for seven different populations. [$\alpha$/Fe]
is defined as the average value of [Mg/Fe], [Si/Fe], [Ca/Fe] and [Ti/Fe],
and may be used to estimate the time-scale for the chemical enrichment 
of the population, as explained in Sect. \ref{sect:intermediate}. 
On the basis of this figure and the more detailed
discussion in the previous sections, the following conclusions
about stellar populations in the Galactic disk, the bulge and 
the halo can be made. 

The disk consists of two main populations, the thin and the thick disk,
which differ in [$\alpha$/Fe] as well as [C/O] and [Eu/Ba]. The thick
disk is formed on a relatively short time-scale with enrichment from type II SNe only
up to [Fe/H]\,$\simeq -0.4$. Thin-disk stars have lower 
values of [$\alpha$/Fe] due to type Ia SNe contributions to
the chemical enrichment; hence the time-scale of evolution is longer
than in the case of the thick disk. In the metallicity range
$-0.7 <$ [Fe/H] $< -0.4$, a gap in [$\alpha$/Fe] is present
between the two disks. Altogether,
the [$\alpha$/Fe] distribution of disk stars is well explained
by a scenario, for which a period of rapid star formation in the early
Galactic disk was interrupted by a merging satellite galaxy that
`heated' the already formed stars to thick-disk kinematics. This was followed by
a hiatus in star formation, in which metal-poor gas was accreted and
type Ia SNe caused [$\alpha$/Fe] to decrease. When
star formation resumed, the first thin disk stars 
formed with low metallicity and low [$\alpha$/Fe]. However, as discussed in 
Sect. \ref{sect:alphadistr}, an alternative model with a monotonically decreasing
star formation rate  and radial migration of stars and gas 
(Sch\"{o}nrich \& Binney 2009a, 2009b) also predicts a bimodal distribution
of [$\alpha$/Fe]. Very precise measurements of the distributions of 
[$\alpha$/Fe] as well as [C/O] and [Eu/Ba] for a large volume-limited
sample of F and G main-sequence stars would be important to 
distinguish between the two competing models for disk
formation.

For the bulge, there has been great progress in studies of abundance
ratios in recent years. In some works, enhanced values of 
[$\alpha$/Fe]\,$\sim +0.3$ are found for stars as metal-rich as the Sun,
but in the most precise differential studies (Alves-Brito et al. 2010),
the trend of [$\alpha$/Fe] for the bulge is found to be similar to that 
of the thick disk (Fig. \ref{fig:bulge.alpha-fe}).
The abundances derived by Bensby et al. (2011) for microlensed  
main-sequence stars (Fig. \ref{fig:all.alpha-fe}) suggest 
that the bulge may consist of two distinct populations:
i.e. old metal-poor stars with enhanced [$\alpha$/Fe] ratios
related to the thick disk, and younger very metal-rich stars
with disk-like $\alpha$/Fe ratios. Still, one would like to see many more
bulge stars observed before drawing
conclusions concerning models for bulge formation from the
abundance ratios.

For halo stars in the solar neighbourhood, there is evidence 
for the existence of two distinct populations clearly separated in
[$\alpha$/Fe], [Na/Fe] and [Ni/Fe] as discussed in 
Sect. \ref{sect:twohalopop}. The high-$\alpha$ stars have abundance ratios
very similar to thick-disk stars. They may be ancient stars formed 
in the Galactic disk or bulge and ejected to the halo by merging satellite galaxies
(Purcell et al. 2010; Zolotov et al. 2009, 2010),
or they may simply belong to the high-velocity
tail of a thick disk with a non-Gaussian velocity distribution. 
The low-$\alpha$ stars tend to have retrograde motions and many of them
move on elongated radial orbits close to the Galactic plane as predicted
for the stellar debris of the captured $\omega$\,Cen progenitor galaxy.
It is likely that the low-alpha stars have
been accreted from dSph galaxies with a relatively slow chemical evolution,
for which type Ia SNe started to contribute iron at [Fe/H]\,$\sim -1.5$.
Perhaps future precise studies of abundance ratios of larger samples
of halo stars will reveal additional subpopulations.

Stars in surviving dSph galaxies have even lower values of [$\alpha$/Fe]
than the low-$\alpha$ halo stars as seen from Fig. \ref{fig:all.alpha-fe}.
This is to be expected according to simulations of the
chemical evolution of a hierarchically formed stellar halo in a
$\Lambda$CDM Universe (Font et al. 2006). Present day dSph galaxies
in the outer halo have experienced a slower chemical evolution than 
more massive satellite galaxies accreted in the early Galxy.  

All globular clusters seem to consist
of multiple stellar populations characterized by different values
of [O/Fe], [Na/Fe], [Al/Fe], and [Mg/Fe]. This may be due to
chemical enrichment from intermediate-mass AGB stars undergoing
hot-bottom hydrogen burning. There is also evidence
for variations of [Fe/H] and the abundance of $s$-process elements
in several globular clusters, most notable in $\omega$\,Cen.
In this connection, it has been suggested that globular
clusters were once the nuclei of now dissolved dwarf galaxies. 

Precise abundance ratios of field stars belonging to the Galactic
halo have so far only been obtained in a small region around the Sun.
It would be important to extend such studies to more distant
halo regions. To do this in an efficient way one needs 
a fiber-coupled high-resolution spectrograph that would make it
possible to observe many stars simultaneously over a relatively
large field, say one degree in diameter. Such a spectrograph
would also be very useful in exploring abundance gradients in
the Galactic disk, both radially and in the direction 
towards the Galactic poles.

\newpage

{\bf List of abbreviations}

\medskip
\noindent
AGB: Asymptotic Giant Branch \\
CDM: Cold Dark Matter \\
CRIRES: CRyogenic InfraRed Echelle Spectrograph \\
DEIMOS: DEep Imaging Multi-Object Spectrograph \\
DLA: Damped Lyman-Alpha \\
dSph: dwarf Spheroidal \\
ESA: European Space Agency \\
ESO: European Southern Observatory \\
EW: Equivalent Width \\
FIES: FIbre fed Echelle Spectrograph \\
FLAMES: Fibre Large Array Multi-Element Spectrograph \\
GAIA: (ESA mission for exploring the Galaxy) \\
HARPS: High Accuracy Radial velocity Planet Searcher \\
HFS: Hyper-Fine Structure \\
HIRES: HIgh Resolution Echelle Spectrometer \\
H-R: Hertzsprung-Russell \\ 
IMF: Initial Mass Function \\
LMC: Large Magellanic Cloud \\
LSR: Local Standard of Rest \\
LTE: Local Thermodynamic Equilibrium \\
NOT: Nordic Optical Telescope \\
RAVE: Radial Velocity Experiment \\
SDSS: Sloan Digital Sky Survey \\
SNe: Supernovae \\
UVES: Ultra-violet and Visible Echelle Spectrograph \\
VLT: Very Large Telescope \\

\medskip
{\bf Index terms }

\medskip
\noindent
Chemical evolution \\
Dwarf galaxies \\  
Echelle spectrographs \\
Effective temperature  \\
Galactic halo \\
Galactic disk  \\
Galactic bulge  \\
Globular clusters \\
Nucleosynthesis  \\
Stellar atmospheres \\
Stellar abundances  \\
Stellar metallicity \\
Stellar populations \\


\begin{thebibliography}{99.}

\bibitem{alves-brito10}
Alves-Brito, A., Mel\'{e}ndez, J., Asplund, M., Ram\'{\i}rez, I., \& Yong, D. 2010,
A\&A, 513, A35

\bibitem{arlandini99}
Arlandini, C., K{\"a}ppeler, F., Wisshak, K., et al. 1999,
ApJ, 525, 886

\bibitem{arnett71}
Arnett, W. D. 1971,
ApJ, 166, 153

\bibitem{asplund05}
Asplund, M. 2005,
ARA\&A, 43, 481

\bibitem{asplund09}
Asplund, M., Grevesse, N., Sauval, A.J., \& Scott, P. 2009,
ARA\&A, 47, 481
 
\bibitem{barklem05}
Barklem, P. S., Christlieb, N., Beers, T. C., et al. 2005,
A\&A, 439, 129


\bibitem{bekki03}
Bekki, K., \& Freeman, K. C. 2003,
MNRAS, 346, L11

\bibitem{bensby06}
Bensby, T., \& Feltzing, S. 2006,
MNRAS, 367, 1181

\bibitem{bensby11}
Bensby, T., Ad{\'e}n, D., Mel{\'e}dez, J., et al. 2011,
A\&A, 533, A134 

\bibitem{bensby05}
Bensby, T., Feltzing, S., Lundstr{\"o}m, I., \& Ilyin, I. 2005,
A\&A, 433, 185

\bibitem{boeche}
Boeche, C., Siebert, A., \& Steinmetz, M. 2008,
AIP Conference Ser., 1082, 61

\bibitem{bergemann10a}
Bergemann, M., \& Cescutti, G. 2010,
A\&A, 522, A9

\bibitem{bergemann08}
Bergemann, M., \& Gehren, T. 2008,
A\&A, 492, 823

\bibitem{bergemann10b}
Bergemann, M., Pickering, J. C., \& Gehren, T. 2010,
MNRAS, 401, 1334

\bibitem{busso99}
Busso, M., Gallino, R., \& Wasserburg, G. J. 1999,
ARA\&A, 37, 239

\bibitem{carollo07}
Carollo, D., Beers, T. C., Lee, Y. S., et al. 2007,
Nature, 450, 1020

\bibitem{caretta09}
Carretta, E., Bragaglia, A., Gratton, R. G., \& Lucatello, S. 2009,
A\&A, 505, 139

\bibitem{caretta10}
Carretta, E., Bragaglia, A., Gratton, R., et al. 2010,
ApJ, 712, L21

\bibitem{casagrande10}
Casagrande, L., Ram\'{\i}rez, I., Mel{\'e}ndez, J., Bessell, M., \& Asplund, M. 2010,
A\&A, 512, A54

\bibitem{cayrel04}
Cayrel, R., Depagne, E., Spite, M., et al. 2004,
A\&A, 416, 1117

\bibitem{chen02}
Chen, Y. Q., Nissen, P. E., Zhao, G., \& Asplund, A. 2002,
A\&A, 390, 225

\bibitem{chen00}
Chen, Y. Q., Nissen, P. E., Zhao, G., Zhang, H. W., \& Benoni, T. 2000,
A\&AS, 141, 491

\bibitem{chiappini06}
Chiappini, C., Hirschi, R., Meynet, G., et al. 2006,
A\&A, 449, L27


\bibitem{cooke11}
Cooke, R., Pettini, M., Steidel, C.C., Rudie, G.C., \& Nissen, P.E. 2011,
MNRAS (in press), arXiv:1106.2805

\bibitem{cunha02}
Cunha, K., Smith, V. V., Suntzeff, N. B., et al. 2002,
AJ, 124, 379

\bibitem{decressin07}
Decressin, T., Meynet, G., Charbonnel, C., Prantzos, N., \& Ekstr{\"o}m, S. 2007,
A\&A, 464, 1029

\bibitem{dercole08}
D'Ercole, A., Vesperini, E., D'Antona, F., et al. 2008,
MNRAS, 391, 825

\bibitem{drawin69}
Drawin, H. W. 1969,
Z. Phys., 225, 483

\bibitem{edvardsson93}
Edvardsson, B., Andersen, J., Gustafsson, B., et al. 1993,
A\&A, 275, 101

\bibitem{eggen62}
Eggen, O. J., Lynden-Bell, D., \& Sandage, A. R. 1962,
ApJ, 136, 748

\bibitem{elmegreen08}
Elmegreen, B. G., Bournaud, F., \& Elmegreen, D. M. 2008,
ApJ, 688, 67

\bibitem{fabbian09}
Fabbian, D., Nissen, P. E., Asplund, M., Pettini, M., \& Akerman, C. 2009,
A\&A, 500, 1143

\bibitem{fenner06}
Fenner, Y., Gibson, B. K., Gallino, R., \& Lugaro, M. 2006,
ApJ, 646, 184

\bibitem{font06}
Font, A. S., Johnston, K. V., Bullock, J. S., \& Robertson, B. E. 2006,
ApJ, 638, 585

\bibitem{fuhrmann04}
Fuhrmann, K. 2004,
AN, 325, 3

\bibitem{fulbright02}
Fulbright, J. P. 2002,
AJ, 123, 404

\bibitem{fulbright07}
Fulbright, J. P., McWilliam, A., \& Rich, R. M. 2007,
ApJ, 661, 1152

\bibitem{gehren04}
Gehren, T., Liang, Y. C., Shi, J. R., Zhang, H. W., \& Zhao, G. 2004,
A\&A, 413, 1045

\bibitem{gilmore83}
Gilmore, G., \& Reid, N. 1983,
MNRAS, 202, 1025

\bibitem{gonzalez08}
Gonz{\'a}lez Hern{\'a}ndez, J. I., \& Bonifacio, P. 2009,
A\&A, 497, 497

\bibitem{gratton01}
Gratton, R. G., Bonifacio, P., Bragaglia, A., et al. 2001,
A\&A, 369, 87

\bibitem{gratton03}
Gratton, R. G., Carretta, E., Desidera, S., et al. 2003,
A\&A, 406, 131

\bibitem{gratton96}
Gratton, R., Carretta, E., Matteucci, F., \& Sneden, C. 1996,
ASP Conf. Ser., 92, 307

\bibitem{gratton04}
Gratton, R. G., Sneden, S., \& Carretta, E. 2004,
ARA\&A, 42, 385

\bibitem{gustafsson08}
Gustafsson, B., Edvardsson, B., Eriksson, K., et al. 2008,
A\&A, 486, 951

\bibitem{haywood08}
Haywood, M. 2008,
MNRAS, 388, 1175

\bibitem{hill11}
Hill, V., Lecureur, A., G{\'o}mez, A., et al. 2011
A\&A (in press), arXiv:1107.5199

\bibitem{ibata94}
Ibata, R. A., Gilmore, G.,\& Irwin, M. J. 1994,
Nature, 370, 194

\bibitem{israelian01}
Israelian, G., \& Rebolo, R. 2001,
ApJ, 557, L43

\bibitem{kirby09}
Kirby, E. N., Guhathakurta, P., Bolte, M., Sneden, C., \& Geha, M. C. 2009,
ApJ, 705, 328

\bibitem{kormendy04}
Kormendy, J., \& Kennicutt Jr., R. C. 2004,
ARA\&A, 42, 603

\bibitem{korn07}
Korn, A. J., Grundahl, F., Richard, O., et al. 2007,
ApJ, 671, 402

\bibitem{kurucz93}
Kurucz, R. 1993,
ATLAS9 Stellar Atmosphere Programs and 2 km/s grid.
Kurucz CD-ROM No.~13. Cambridge, Mass.,
Smithsonian Astrophysical Observatory

\bibitem{lecureur07}
Lecureur, A., Hill, V., Zoccali, M., et al. 2007,
A\&A, 465, 799

\bibitem{lee09}
Lee, J-W., Kang, Y-W., Lee, J., \& Lee, Y-W. 2009,
Nature, 462, 480

\bibitem{lee11}
Lee, Y.S., Beers, T.C., An, D., et al. 2011, 
ApJ, 738, 187

\bibitem{luck06}
Luck, R. E., Kovtyukh, V. V., \& Andrievsky, S. M. 2006,
AJ, 132, 902

\bibitem{marino09}
Marino, A. F., Milone, A. P., Piotto, G., et al. 2009,
A\&A, 505, 1099

\bibitem{mashonkina11}
Mashonkina, L., Gehren, T., Shi, J.-R., Korn, A.J., \& Grupp, F. 2011,
A\&A, 528, A87

\bibitem{mashonkina03}
Mashonkina, L., Gehren, T., Travaglio, C., \& Borkova, T. 2003,
A\&A, 397, 275

\bibitem{mcwilliam95}
McWilliam, A., Preston, G. W., Sneden, C., \& Searle, L. 1995,
AJ, 109, 2757

\bibitem{mcwilliam94}
McWilliam, A., \& Rich, R. M. 1994,
ApJS, 91, 749

\bibitem{melendez08}
Mel{\'e}ndez, J., Asplund, M., Alves-Brito, A., et al. 2008,
A\&A, 484, L21

\bibitem{melendez09}
Mel{\'e}ndez, J., Asplund, M., Gustafsson, B., \& Yong, D. 2009,
ApJ, 704, L66

\bibitem{meza05}
Meza, A., Navarro, J. F., Abadi, M. G., \& Steinmetz, M. 2005,
MNRAS, 359, 93
 
\bibitem{mishenina02}
Mishenina, T. V., Kovtyukh, V. V., Soubiran, C., Travaglio, C., \& Busso, M. 2002,
A\&A, 396, 189
 
\bibitem{neves09}
Neves, V., Santos, N. C., Sousa, S. G., Correia, A. C. M., \& Israelian, G. 2009,
A\&A, 497, 563

\bibitem{nissen08}
Nissen, P. E. 2008,
Physica Scripta, T133, 014022 

\bibitem{nissen07}
Nissen, P. E., Akerman, C., Asplund, M., et al. 2007,
A\&A, 469, 319


\bibitem{nissen02}
Nissen, P. E., Primas, F., Asplund, M., \& Lambert, D. L. 2002,
A\&A, 390, 235

\bibitem{nissen97}
Nissen, P. E., \& Schuster, W. J. 1997,
A\&A, 326, 751


\bibitem{nissen10}
Nissen, P. E., \& Schuster, W. J. 2010,
A\&A, 511, L10 

\bibitem{nissen11}
Nissen, P. E., \& Schuster, W. J. 2011,
A\&A, 530, A15 

\bibitem{nordstrom04}
Nordstr{\"o}m, B., Mayor, M., Andersen, J., et al. 2004,
A\&A, 418, 989



\bibitem{pompeia08}
Pomp{\'e}ia, L., Hill, V., Spite, M., et al. 2008,
A\&A, 480, 379

\bibitem{purcell10}
Purcell, C. W., Bullock, J. S., \& Kazantzidis, S. 2010,
MNRAS, 404, 1711

\bibitem{ramirez07}
Ram{\'{\i}}rez, I., Allende Prieto, C., \& Lambert, D. L. 2007,
A\&A, 465, 271

\bibitem{reddy06}
Reddy, B. E., Lambert, D. L., \& Allende Prieto, C. 2006,
MNRAS, 367, 1329

\bibitem{reddy03}
Reddy, B. E., Tomkin, J., Lambert, D. L., \& Allende Prieto, C. 2003,
MNRAS, 340, 304

\bibitem{romano07}
Romano, D., \& Matteucci, F. 2007,
MNRAS, 378, L59

\bibitem{ryde10}
Ryde, N., Gustafsson, B., Edvardsson, B., et al. 2010,
A\&A, 509, A26

\bibitem{sbordone07}
Sbordone, L., Bonifacio, P., Buonanno, R., et al. 2007,
A\&A, 465, 815

\bibitem{schonrich09a}
Sch\"{o}nrich, R., \& Binney, J. 2009a,
MNRAS, 396, 203

\bibitem{schonrich09b}
Sch\"{o}nrich, R., \& Binney, J. 2009b,
MNRAS, 399, 1145

\bibitem{schuster06}
Schuster, W. J., Moitinho, A., M\'{a}rquez, A., Parrao, L., \& Covarrubias, E. 2006,
A\&A, 445, 939

\bibitem{searle78}
Searle, L., \& Zinn, R. 1978,
ApJ, 225, 357

\bibitem{smith00}
Smith, V. V., Suntzeff, N. B., Cunha, K., et al. 2000,
AJ, 119, 1239

\bibitem{sousa07}
Sousa, S. G., Santos, N. C., Israelian, G., Mayor, M., \& Monteiro, M. J. P. F. G. 2007,
A\&A, 469, 783 

\bibitem{stephens02}
Stephens, A., \& Boesgaard, A. M. 2002,
AJ, 123, 1647

\bibitem{stromgren87}
Str{\"o}mgren, B. 1987,
In The Galaxy, ed. G. Gilmore \& B. Carswell (Reidel, Dordrecht), 229

\bibitem{takeda05}
Takeda, Y., Hashimoto, O., \& Taguchi, H. 2005,
PASJ, 57, 751

\bibitem{tolstoy09}
Tolstoy, E., Hill, V., \& Tosi, M. 2009, 
ARA\&A, 47, 371

\bibitem{tsujimoto95}
Tsujimoto, T., Nomoto, K., Yoshii, Y., et al. 1995,
MNRAS, 277, 945

\bibitem{turcotte02}
Turcotte, S., \& Wimmer-Schweingruber, R. F. 2002,
Journal of Geophysical Research, 107, 1442

\bibitem{venn04}
Venn, K. A., Irwin, M., Shetrone, M. D., et al. 2004,
AJ, 128, 1177

\bibitem{ventura01}
Ventura, P., D'Antona, F., Mazzitelli, I., \& Gratton, R. 2001,
ApJ, 550, L65

\bibitem{yong06}
Yong, D., Carney, B. W., Teixera de Almeida, M. L., \& Poha, B. L. 2006,
AJ, 131, 2256

\bibitem{yong08}
Yong, D., \& Grundahl, F. 2008,
ApJ, 672, L29

\bibitem{yong03}
Yong, D., Grundahl, F., Lambert, D. L., Nissen, P. E., \& Shetrone, M. D. 2003,
A\&A, 402, 985

\bibitem{yong05}
Yong, D., Grundahl, F., Nissen, P. E., Jensen, H. R., \& Lambert, D. L. 2005,
A\&A, 438, 875

\bibitem{zoccali08}
Zoccali, M., Hill, V., Lecureur, A., et al. 2008,
A\&A, 486, 177

\bibitem{zolotov09}
Zolotov, A., Willman, B., Brooks, A. M., et al. 2009,
ApJ, 702, 1058

\bibitem{zolotov10}
Zolotov, A., Willman, B., Brooks, A. M., et al. 2010,
ApJ, 721, 738

\end{thebibliography}
\end{document}